\PassOptionsToPackage{unicode}{hyperref}
\PassOptionsToPackage{hyphens}{url}
\PassOptionsToPackage{dvipsnames,svgnames,x11names}{xcolor}
\documentclass[
  10pt,
  letterpaper,
]{article}
\usepackage{xcolor}
\usepackage[margin=0.78in]{geometry}
\usepackage{amsmath,amssymb}
\usepackage{iftex}
\ifPDFTeX
  \usepackage[T1]{fontenc}
  \usepackage[utf8]{inputenc}
  \usepackage{textcomp} 
\else 
  \usepackage{unicode-math} 
  \defaultfontfeatures{Scale=MatchLowercase}
  \defaultfontfeatures[\rmfamily]{Ligatures=TeX,Scale=1}
\fi
\usepackage{lmodern}
\ifPDFTeX\else
\fi
\IfFileExists{upquote.sty}{\usepackage{upquote}}{}
\IfFileExists{microtype.sty}{
  \usepackage[]{microtype}
  \UseMicrotypeSet[protrusion]{basicmath} 
}{}
\usepackage{setspace}
\makeatletter
\@ifundefined{KOMAClassName}{
  \IfFileExists{parskip.sty}{%
    \usepackage{parskip}
  }{
    \setlength{\parindent}{0pt}
    \setlength{\parskip}{6pt plus 2pt minus 1pt}}
}{
  \KOMAoptions{parskip=half}}
\makeatother
\usepackage{color}
\usepackage{fancyvrb}

\DefineVerbatimEnvironment{Highlighting}{Verbatim}{commandchars=\\\{\}}
\newenvironment{Shaded}{}{}

\newcommand{\AttributeTok}[1]{\textcolor[rgb]{0.49,0.56,0.16}{#1}}

\newcommand{\ExtensionTok}[1]{#1}

\newcommand{\NormalTok}[1]{#1}

\usepackage{longtable,booktabs,array}
\usepackage{calc} 
\usepackage{etoolbox}
\makeatletter
\patchcmd\longtable{\par}{\if@noskipsec\mbox{}\fi\par}{}{}
\makeatother
\IfFileExists{footnotehyper.sty}{\usepackage{footnotehyper}}{\usepackage{footnote}}
\makesavenoteenv{longtable}
\usepackage{graphicx}
\makeatletter
\newsavebox\pandoc@box
\newcommand*\pandocbounded[1]{
  \sbox\pandoc@box{#1}%
  \Gscale@div\@tempa{\textheight}{\dimexpr\ht\pandoc@box+\dp\pandoc@box\relax}%
  \Gscale@div\@tempb{\linewidth}{\wd\pandoc@box}%
  \ifdim\@tempb\p@<\@tempa\p@\let\@tempa\@tempb\fi
  \ifdim\@tempa\p@<\p@\scalebox{\@tempa}{\usebox\pandoc@box}%
  \else\usebox{\pandoc@box}%
  \fi%
}
\def\fps@figure{htbp}
\makeatother
\providecommand{\tightlist}{%
  \setlength{\itemsep}{0pt}\setlength{\parskip}{0pt}}
\usepackage{bookmark}
\IfFileExists{xurl.sty}{\usepackage{xurl}}{} 
\hypersetup{
  pdftitle={Fail-Closed Lowering of Resident KV Claims onto LLM Serving Runtimes},
  pdfauthor={Lukas Stepanek luki.step@proton.me},
  colorlinks=true,
  linkcolor={blue},
  filecolor={Maroon},
  citecolor={Blue},
  urlcolor={blue},
  pdfcreator={LaTeX via pandoc}}

\title{Fail-Closed Lowering of Resident KV Claims onto LLM Serving
Runtimes}
\author{Lukas Stepanek\\
\href{mailto:luki.step@proton.me}{\nolinkurl{luki.step@proton.me}}}
\date{May 2026}

\usepackage{microtype}
\usepackage{booktabs}
\usepackage{array}
\usepackage{float}
\usepackage{xurl}
\begin{document}
\maketitle

\setstretch{1.03}
\section{Abstract}\label{abstract}

LLM serving runtimes increasingly expose KV-cache primitives that look
like future-reuse controls: retention priority, TTL-like duration, host
or storage offload, block events, active no-evict scheduling, and
KV-aware routing. Feature primitives are weaker than accepted future-KV
obligations: a runtime can expose priority, offload, events, and routing
without accepting responsibility for a future reuse claim. This paper
studies ResidentClaim lowering at the obligation level. A runtime or
trusted adapter must bind behavior to accepted claim identity, a
materialization predicate, ordered lifecycle events, and claim-scoped
outcomes before a primitive can be treated as semantic conformance. We
contribute a fail-closed lowering relation, checker, descriptor format,
and bad-lowering suite that classify rows as native conformance,
adapter-observational evidence, adapter-policy evidence under controlled
pressure, approximation substrate, rejected mapping, or unknown
evidence. In the descriptors and evidence audited here, the checker
validates manually curated, anchored runtime descriptors against
obligation bundles; it is not a proof that unaudited runtime behavior is
complete. Public TensorRT-LLM, SGLang/HiCache, and Dynamo expose strong
substrates and selected adapter positives, not native ResidentClaim
conformance. The positive systems witness is a local patched vLLM
connector/scheduler-boundary mechanism: claim metadata flows through
real in-process offload/load behavior, and controlled same-claim
restoration failure reaches vLLM's invalid-KV-load path and becomes an
ordered scheduler-boundary fail-closed active outcome. The result is a
calibrated semantics boundary, not a production performance claim or a
compatibility survey.

\section{1. Introduction}\label{introduction}

KV-cache reuse has become an explicit systems surface in LLM serving.
Production and research runtimes expose token-range retention
priorities, duration fields, block stored/removed events,
GPU/host/storage cache tiers, load-back paths, active no-evict modes,
and KV-aware routing. These mechanisms are real and important. They also
create a tempting inference: if a runtime has priority, offload,
routing, or no-evict, perhaps it implements future-resident KV claims.

This paper argues that the inference is unsound. A ResidentClaim is not
a knob name. It is an accepted future-reuse responsibility over a cache
identity, a reusable object, a materialization predicate, a footprint, a
mode, and an ordered outcome. If a block is stored, that does not show
that the claimed future computation state is materialized. If a block is
removed, that does not show claim harm. If bytes move to host memory,
that does not show the claim can be restored before reuse. If a router
chooses a worker with useful cached KV, that does not show the route and
later reuse are attributed to an accepted claim.

The checker validates curated descriptors and anchored evidence against
these obligation bundles. It deliberately does not prove that unaudited
runtime behavior is complete. This distinction is the main calibration
point: feature names can be useful substrates, but accepted future-reuse
obligations require claim identity, acceptance, predicate, order, and
claim-scoped outcomes at the depth being asserted.

We treat ResidentClaims as the contract abstraction and focus on the
lowering question: when can a runtime primitive, adapter, or patch
conformantly supply the obligations of that contract? This keeps the
paper's role separate from the contract definition itself. The
contribution here is the evidence relation that decides whether concrete
serving surfaces satisfy, approximate, or fail those obligations.

The first claim is semantic: accepted future-KV responsibility requires
claim-level obligations. Without claim identity, a runtime cannot
distinguish ordinary cache behavior from claim outcomes. Without
explicit acceptance, there is no responsibility boundary. Without a
predicate, block survival is not useful state survival. Without ordered
lifecycle events, an observer cannot distinguish restore-before-reuse
from restore-after-use, or demotion-before-loss from post-hoc
explanation. Without failure outcomes, an offload substrate does not
become an offloadable claim.

A running example fixes the boundary. Let claim \texttt{C} cover prefix
\texttt{P} with predicate \texttt{leading-\allowbreak{}prefix-\allowbreak{}at-\allowbreak{}least-\allowbreak{}448}. The
claim is first accepted by a mechanism that records \texttt{C},
\texttt{P}, and the predicate. The prefix becomes materialized; later,
its KV blocks are offloaded from GPU to CPU. A future reuse request for
\texttt{P} is not satisfied merely because CPU bytes exist. It requires
an ordered CPU-to-GPU restoration before the reuse consumes the prefix.
In the patched vLLM connector/scheduler-boundary mechanism, the failure
experiment forces that same-claim restoration to fail. The scheduler
invalid-KV-load path then reports a claim-scoped restoration failure and
a fail-closed active outcome for \texttt{C}. This is the event sequence
the checker accepts; generic transfer counters, fallback recomputation,
wrong-claim failure, or unclaimed failure are not accepted as satisfying
\texttt{C}.

For this example, the accepted path is deliberately narrow:
\texttt{accept(C,\ P,\ leading-\allowbreak{}prefix-\allowbreak{}at-\allowbreak{}least-\allowbreak{}448)},
\texttt{materialized(C)}, \texttt{offloaded(C)},
\texttt{restore-\allowbreak{}required(C)}, same-claim load failure,
\texttt{restoration\_\allowbreak{}failed(C)}, and then
\texttt{active\_request\_refused(blocking\_claim\_ids={[}C{]})}. The
same bytes, counters, or request error without that ordering are
insufficient.

The second claim is methodological: a fail-closed lowering relation can
separate native conformance, adapter-scoped evidence, approximation
substrates, rejected mappings, and unknown evidence. This matters
because several rows are partly positive. TensorRT-LLM block events can
be joined by an external adapter to reconstruct selected
\texttt{best\_\allowbreak{}effort} materialization observations. TensorRT priority
can be joined to external claim state to reconstruct controlled
soft-priority retention-order evidence. SGLang request cached-token
telemetry and block events can be joined to reconstruct selected
materialization observations. These rows are useful, but they are not
native backend support and they do not imply enforcement.

The third claim is empirical and systems-oriented: public runtimes
expose strong substrates and narrow adapter positives, while a targeted
patched runtime mechanism demonstrates implementability of the missing
lifecycle/outcome semantics. In the current artifacts, no TensorRT-LLM,
SGLang/HiCache, or Dynamo descriptor produces \texttt{native\_\allowbreak{}sound}.
The local patched vLLM connector/scheduler-boundary mechanism does pass
the offload lifecycle/outcome gate at \texttt{backend\_\allowbreak{}patch} depth:
claim metadata flows through real in-process
\texttt{Offloading\allowbreak{}Connector} store/load jobs and worker transfers, and
controlled same-claim \texttt{CPU\ -\textgreater{}\ GPU} load failure
produces \texttt{scheduler\_\allowbreak{}resident\_\allowbreak{}claim\_\allowbreak{}restoration\_\allowbreak{}failed}
followed by the fail-closed active outcome event
\texttt{scheduler\_\allowbreak{}active\_\allowbreak{}request\_\allowbreak{}refused} before request termination.

Negative rows are a result, not a weakness in the evaluation. A
conventional compatibility table would mark ``priority'', ``TTL'',
``offload'', ``events'', and ``routing'' and appear to close the
problem. The lowering matrix asks a different question: which
obligations are actually represented, at which adapter depth, under
which evidence scope? When active no-evict is active-request scoped
rather than accepted future-resident protection, the correct result is
rejection. When storage tiers lack restoration-before-reuse and
restoration-failure outcomes, the correct result is approximation. When
routing lacks claim-scoped route cost, placement, and reuse attribution,
the correct result is approximation. Those classifications are what keep
the contract meaningful.

\textbf{Intro counterexamples: feature names versus obligations.}

{\def\LTcaptype{none} 
\begin{longtable}[]{@{}
  >{\raggedright\arraybackslash}p{(\linewidth - 4\tabcolsep) * \real{0.3333}}
  >{\raggedright\arraybackslash}p{(\linewidth - 4\tabcolsep) * \real{0.3333}}
  >{\raggedright\arraybackslash}p{(\linewidth - 4\tabcolsep) * \real{0.3333}}@{}}
\toprule\noalign{}
\begin{minipage}[b]{\linewidth}\raggedright
Feature-only inference
\end{minipage} & \begin{minipage}[b]{\linewidth}\raggedright
Missing obligation boundary
\end{minipage} & \begin{minipage}[b]{\linewidth}\raggedright
Correct lowering label
\end{minipage} \\
\midrule\noalign{}
\endhead
\bottomrule\noalign{}
\endlastfoot
Priority value or retention hint means \texttt{soft\_\allowbreak{}priority}. & Claim
identity, claim-scoped telemetry, and controlled evidence that priority
changed retention order. & \texttt{approximate} unless the pressure and
claim-join evidence are present. \\
Duration/TTL field means \texttt{expiring}. & Claim-scoped expiry
boundary and ordered responsibility transition before loss. &
\texttt{approximate} under current public evidence. \\
Active no-evict means future-resident \texttt{hard\_\allowbreak{}protected}. &
Accepted future-resident identity, protected-victim exclusion, explicit
conflict action, blocking claim ids, harm attribution, and order. &
\texttt{rejected} for the observed active-request-scoped mapping. \\
Host/storage tier means \texttt{offloadable}. & Restoration before
predicate-satisfying reuse and claim-scoped restoration failure outcome.
& \texttt{approximate} or \texttt{unknown}, depending on whether the
storage substrate is exercised. \\
KV-aware routing means \texttt{routed\_\allowbreak{}reuse}. & Claim-scoped route
cost, placement attribution, later reuse attribution, predicate, and
telemetry. & \texttt{approximate} under current docs-backed evidence. \\
Generic transfer or request-failure counters mean restoration failure. &
Same accepted claim, ordered restore-required event, scheduler-boundary
failure/refusal attribution, and blocking ids. & Rejected by the
connector gate. \\
\end{longtable}
}

This paper is not a KV eviction algorithm, not a production offload
evaluation, and not a claim that existing runtimes lack useful KV
mechanisms. It is a semantic conformance study: what must be true before
an accepted ResidentClaim can be lowered onto a serving runtime, and
what current evidence can and cannot show.

The paper is therefore organized around three pillars: obligation-level
lowering and a fail-closed checker; public-runtime boundary studies with
false-positive rejection; and the patched vLLM
connector/scheduler-boundary mechanism as the positive systems witness
for the missing offload lifecycle/outcome semantics.

\section{2. Contributions}\label{contributions}

This paper makes four concrete contributions.

\begin{enumerate}
\def\labelenumi{\arabic{enumi}.}
\item
  It defines an obligation-based lowering model for ResidentClaim modes:
  \texttt{best\_\allowbreak{}effort}, \texttt{soft\_\allowbreak{}priority},
  \texttt{hard\_\allowbreak{}protected}, \texttt{demotable}, \texttt{expiring},
  \texttt{offloadable}, and \texttt{routed\_\allowbreak{}reuse}.
\item
  It implements a fail-closed checker and false-positive suite over
  machine-readable runtime descriptors. The checker counts obligations
  only when evidence is supported and anchored, requires anchored
  observed evidence atoms, and keeps adapter depth separate from
  classification. Descriptor and evidence mutation controls fail closed
  in 16/16 cases.
\item
  It studies the boundary of public TensorRT-LLM, SGLang/HiCache,
  Dynamo-style KV routing, and vLLM surfaces. The resulting matrix
  distinguishes adapter-observational evidence, adapter-policy evidence
  under controlled pressure, approximation substrates, rejected
  lowerings, and unknown rows.
\item
  It demonstrates a local patched vLLM connector/scheduler-boundary
  mechanism at \texttt{backend\_\allowbreak{}patch} depth. The repeated
  scheduler-boundary evaluation records 131/131 completed subprocesses,
  131/131 valid event sequences, 30/30 successful observation passes,
  30/30 same-claim scheduler-boundary failure-outcome passes, and
  fail-closed rejection of wrong-claim, unclaimed, fallback-recompute,
  ordinary-offload-without-claim, and generic-counter controls. A clean
  multi-claim attribution rerun records 3/3 target-only failure/refusal
  attributions while the non-target claim restores successfully.
\end{enumerate}

The contribution is deliberately bounded. We do not claim a better KV
eviction policy, production performance, native/upstream vLLM support,
native TensorRT-LLM/SGLang/Dynamo support, production offloadable
support, pre-admission refusal, or scheduler-native admission protocol
support.

\section{3. ResidentClaim Obligations}\label{residentclaim-obligations}

ResidentClaim lowering is strict because accepted responsibility is
strict. The obligations are not arbitrary hurdles for existing systems;
they define a conservative audit boundary for deciding whether a
future-reuse claim was satisfied, demoted, expired, restored, refused,
harmed, or simply never accepted. Each obligation blocks a common false
positive where ordinary cache behavior is misread as accepted-claim
conformance. This is not a proven minimality claim. This section
summarizes the contract obligations only to fix the checker input; the
paper's new object is the lowering judgment over runtime descriptors and
evidence.

\textbf{Table 1: ResidentClaim obligations.}

{\def\LTcaptype{none} 
\begin{longtable}[]{@{}
  >{\raggedright\arraybackslash}p{(\linewidth - 2\tabcolsep) * \real{0.5000}}
  >{\raggedright\arraybackslash}p{(\linewidth - 2\tabcolsep) * \real{0.5000}}@{}}
\toprule\noalign{}
\begin{minipage}[b]{\linewidth}\raggedright
Obligation
\end{minipage} & \begin{minipage}[b]{\linewidth}\raggedright
Why it is necessary
\end{minipage} \\
\midrule\noalign{}
\endhead
\bottomrule\noalign{}
\endlastfoot
\texttt{claim\_\allowbreak{}identity} & Without a stable claim id, block/request
events cannot distinguish claimed KV from unclaimed KV. \\
\texttt{explicit\_\allowbreak{}acceptance} & Without acceptance, an application hint
never becomes runtime responsibility. \\
\texttt{materialization\_\allowbreak{}predicate} & Without a useful-state predicate,
stored blocks cannot be interpreted as satisfying the claimed future
reuse. \\
\texttt{footprint\_\allowbreak{}accounting} & Without footprint at claim granularity,
active/resident infeasibility cannot be attributed or budgeted. \\
\texttt{ordered\_\allowbreak{}lifecycle\_\allowbreak{}events} & Without order, the observer cannot
tell whether demotion, expiry, restore, refusal, or loss happened before
or after the relevant claim transition. \\
\texttt{claim\_\allowbreak{}materialized\_\allowbreak{}event} & Without a claim-scoped
materialization event, best-effort telemetry remains block/request
telemetry. \\
\texttt{claim\_\allowbreak{}demoted\_\allowbreak{}before\_\allowbreak{}loss} & Without an ordered demotion,
losing protected state is indistinguishable from breaking a
still-protected claim. \\
\texttt{claim\_\allowbreak{}expired\_\allowbreak{}boundary} & Without a responsibility boundary,
post-duration loss cannot be classified as violation or post-expiry
non-responsibility. \\
\texttt{offload\_\allowbreak{}restorability} & Without restoration before reuse, a
storage tier is only bytes somewhere else. \\
\texttt{restoration\_\allowbreak{}failure\_\allowbreak{}outcome} & Without claim-scoped failure,
offload failure is indistinguishable from ordinary miss, unrelated
error, or fallback recomputation. \\
\texttt{victim\_\allowbreak{}exclusion\_\allowbreak{}before\_\allowbreak{}violation} & Without victim
exclusion, hard protection collapses to soft priority under pressure
unless another explicit contract transition occurs first. \\
\texttt{explicit\_\allowbreak{}conflict\_\allowbreak{}action} & Explicit active/resident conflict
action: refusal, defer, route elsewhere, bound active KV, offload,
split/recompute, or explicit claim relaxation/demotion when the contract
permits it. \\
\texttt{blocking\_\allowbreak{}claim\_\allowbreak{}ids} & Without blocking ids, the chosen
conflict action cannot be attributed to the resident claim that caused
infeasibility. \\
\texttt{claim\_\allowbreak{}harm\_\allowbreak{}attribution} & Without harm attribution,
predicate-breaking loss is ordinary cache loss, not claim harm. \\
\texttt{claim\_\allowbreak{}scoped\_\allowbreak{}telemetry} & Without claim-scoped telemetry,
observations cannot become claim outcomes. \\
\texttt{priority\_\allowbreak{}influence} & Soft priority requires evidence that
priority affects policy, not merely that a field exists. \\
\texttt{route\_\allowbreak{}cost\_\allowbreak{}attribution} & Without route cost attribution,
routing cannot report what cost was paid to preserve or reuse a
claim. \\
\texttt{placement\_\allowbreak{}attribution} & Without placement attribution, worker
choice is orchestration state, not a claim action. \\
\texttt{reuse\_\allowbreak{}routing\_\allowbreak{}attribution} & Without reuse attribution, later
hit/miss behavior cannot be assigned to the routed claim path. \\
\end{longtable}
}

The checker keeps the older key \texttt{active\_\allowbreak{}refusal\_\allowbreak{}or\_\allowbreak{}defer} only
as a backward-compatible alias for \texttt{explicit\_\allowbreak{}conflict\_\allowbreak{}action};
the manuscript tables and descriptions use the latter name.

The current checker groups these obligations into mode bundles. Each
bundle is defined in \texttt{modes.\allowbreak{}yaml}; the checker test used here is:

\begin{itemize}
\tightlist
\item
  \texttt{best\_\allowbreak{}effort}: claim identity, a fixed materialization
  predicate, and claim-scoped materialization telemetry. Raw
  stored/removed block events are not enough.
\item
  \texttt{soft\_\allowbreak{}priority}: claim identity, priority influence,
  claim-scoped telemetry, and anchored pressure observations. Priority
  values or duration fields alone are not enough.
\item
  \texttt{hard\_\allowbreak{}protected}: explicit accepted-claim responsibility,
  footprint accounting, victim exclusion before violation, an explicit
  conflict action, blocking claim ids, harm attribution, and order.
  Refusal/defer is one valid implementation path, not the only one:
  routing elsewhere, bounding active KV, offloading active or resident
  state, recompute/split, or explicit relaxation/demotion can satisfy
  the action-space obligation only if the descriptor claims that action
  and provides anchored evidence for it. Active no-evict for running
  requests is not enough.
\item
  \texttt{demotable} and \texttt{expiring}: an ordered claim-level
  responsibility boundary before loss. Priority updates, lower cache
  levels, or wall-clock arithmetic are not enough.
\item
  \texttt{offloadable}: restoration before predicate-satisfying reuse
  and a claim-scoped restoration-failure outcome. Storage tiers,
  prefetch, write-back, load-back, transfer counters, or generic
  failures are not enough.
\item
  \texttt{routed\_\allowbreak{}reuse}: claim-scoped route cost, placement, and later
  reuse attribution. KV-aware routing or worker overlap scoring alone is
  not enough.
\end{itemize}

\textbf{Evidence-to-obligation summary.}

{\def\LTcaptype{none} 
\begin{longtable}[]{@{}
  >{\raggedright\arraybackslash}p{(\linewidth - 10\tabcolsep) * \real{0.1667}}
  >{\raggedright\arraybackslash}p{(\linewidth - 10\tabcolsep) * \real{0.1667}}
  >{\raggedright\arraybackslash}p{(\linewidth - 10\tabcolsep) * \real{0.1667}}
  >{\raggedright\arraybackslash}p{(\linewidth - 10\tabcolsep) * \real{0.1667}}
  >{\raggedright\arraybackslash}p{(\linewidth - 10\tabcolsep) * \real{0.1667}}
  >{\raggedright\arraybackslash}p{(\linewidth - 10\tabcolsep) * \real{0.1667}}@{}}
\toprule\noalign{}
\begin{minipage}[b]{\linewidth}\raggedright
Mode
\end{minipage} & \begin{minipage}[b]{\linewidth}\raggedright
Required obligations
\end{minipage} & \begin{minipage}[b]{\linewidth}\raggedright
Allowed adapter/patch depth
\end{minipage} & \begin{minipage}[b]{\linewidth}\raggedright
Accepted evidence shape
\end{minipage} & \begin{minipage}[b]{\linewidth}\raggedright
Rejected false positive
\end{minipage} & \begin{minipage}[b]{\linewidth}\raggedright
Evidence source class
\end{minipage} \\
\midrule\noalign{}
\endhead
\bottomrule\noalign{}
\endlastfoot
\texttt{best\_\allowbreak{}effort} & Identity, predicate, materialized event, claim
telemetry. & \texttt{telemetry\_\allowbreak{}join} or deeper under registry/join
preconditions. & Anchored block, token, or request evidence joined to a
pre-registered claim and named predicate at a named observation point. &
Raw block stored/removed events. & Source, trace, artifact-generated. \\
\texttt{soft\_\allowbreak{}priority} & Identity, priority influence, claim telemetry,
and controlled pressure atoms. & Native priority plus
\texttt{telemetry\_\allowbreak{}join} for claim attribution. & Original,
swapped-priority, and equal-priority pressure controls with claim-scoped
retention order. & Priority value, duration field, or source-level
priority API alone. & Source, trace, controlled pressure. \\
\texttt{hard\_\allowbreak{}protected} & Identity, acceptance, predicate, footprint,
victim exclusion, explicit conflict action, blocking ids, harm
attribution, and order. & \texttt{scheduler\_\allowbreak{}hook} plus
\texttt{allocator\_\allowbreak{}hook}, or \texttt{backend\_\allowbreak{}patch} when the patch
supplies both. & Anchored conflict trace for the claimed action:
refusal, defer, route elsewhere, bound active KV, offload,
split/recompute, or explicit relaxation/demotion. & Active no-evict for
the running request. & Source, trace, failure injection,
artifact-generated. \\
\texttt{demotable} & Identity, acceptance, demotion before
predicate-breaking loss, and order. & \texttt{claim\_\allowbreak{}registry} plus
ordered lifecycle evidence, or \texttt{backend\_\allowbreak{}patch}. & Claim-scoped
demotion boundary before loss. & Priority update, lower cache level, or
post-hoc explanation. & Source, trace, artifact-generated. \\
\texttt{expiring} & Identity, acceptance, expiry boundary, and order. &
\texttt{claim\_\allowbreak{}registry} plus ordered lifecycle evidence, or
\texttt{backend\_\allowbreak{}patch}. & Claim-scoped boundary where responsibility
ends before later loss. & TTL/duration metadata alone. & Docs, source,
trace. \\
\texttt{offloadable} & Identity, acceptance, predicate, restorability,
restoration-failure outcome, order, and harm attribution. &
\texttt{storage\_\allowbreak{}restorability} or \texttt{backend\_\allowbreak{}patch}. &
Restore-before-reuse success evidence plus claim-scoped restoration
failure/refusal, demotion, expiry, or harm when restoration fails. &
Storage tier, generic transfer counters, fallback recompute, wrong-claim
failure. & Source, trace, failure injection, artifact-generated. \\
\texttt{routed\_\allowbreak{}reuse} & Identity, predicate, route cost, placement,
reuse attribution, and telemetry. & \texttt{routing\_\allowbreak{}hook} or
\texttt{backend\_\allowbreak{}patch}. & Route decision, placement, and later reuse
success/failure attributed to the accepted claim. & KV-overlap routing
score or worker choice alone. & Docs, source, trace. \\
\end{longtable}
}

Three kinds of ``acceptance'' must also be kept separate. Claim
registration by an external adapter can be sufficient for selected
observational rows if the claim is pre-registered and the join is
unambiguous. Backend primitive acceptance, such as accepting a
retention-priority field, shows only that a primitive was configured.
ResidentClaim semantic acceptance requires enough obligations to treat
the claim as runtime responsibility.

\section{4. Lowering Relation and
Checker}\label{lowering-relation-and-checker}

The core judgment is:

\begin{Shaded}
\begin{Highlighting}[]
\NormalTok{backend + adapter + evidence |= ResidentClaim mode}
\end{Highlighting}
\end{Shaded}

A backend/adapter/evidence tuple lowers a mode only if every required
obligation for that mode is represented by the native backend or by an
adapter whose depth and preconditions allow it to supply that
obligation. Missing required obligations fail closed. The row becomes
\texttt{approximate}, \texttt{rejected}, or \texttt{unknown}, depending
on whether approximation signals exist, enforcement critical obligations
are absent, or the evidence is inconclusive.

The checker core is intentionally small:

\begin{Shaded}
\begin{Highlighting}[]
\NormalTok{O[m]        = the obligation bundle required by mode m}
\NormalTok{d           = descriptor with native evidence d.native}
\NormalTok{a           = selected adapter depth and adapter evidence d.adapters[\textless{}= a]}
\NormalTok{E           = anchored evidence atoms from docs, source, traces, controls}
\NormalTok{e           = one native or adapter evidence item}

\NormalTok{supports(e, o)}
\NormalTok{  true when e marks obligation o as supported and has a concrete anchor.}

\NormalTok{anchored(e)}
\NormalTok{  true when e names an anchor kind, path or public source, and note. For trace}
\NormalTok{  evidence, the anchor must also preserve the relevant order and claim scope.}

\NormalTok{depth\_allowed(a, o)}
\NormalTok{  true when o is native, or adapter depth a is allowed to supply o and its}
\NormalTok{  required preconditions hold.}

\NormalTok{Lower(d, a, E, m)}
\NormalTok{  iff for every o in O[m], there exists evidence e in d.native, d.adapters[\textless{}= a],}
\NormalTok{  or E such that supports(e, o), anchored(e), and depth\_allowed(a, o).}

\NormalTok{label(d, a, E, m):}
\NormalTok{  native\_sound        if Lower holds using native evidence only.}
\NormalTok{  sound\_with\_adapter  if Lower holds using allowed adapter or backend{-}patch}
\NormalTok{                      evidence under its stated trust boundary.}
\NormalTok{  rejected            if the proposed mapping violates a forbidden lowering or}
\NormalTok{                      misses enforcement{-}critical obligations.}
\NormalTok{  approximate         if related primitives or approximation signals exist, but}
\NormalTok{                      Lower does not hold.}
\NormalTok{  unknown             if the evidence is inconclusive and no recognized}
\NormalTok{                      approximation signal is exercised.}
\end{Highlighting}
\end{Shaded}

The checker does not require the exact event names used by the local
vLLM patch. Equivalent runtime evidence can satisfy an obligation if it
is anchored, ordered, claim-scoped, and supplied at an allowed adapter
depth. The patch event names are one concrete witness, not a naming
convention that public runtimes must copy.

The implementation is a small Python checker over YAML mode definitions
and backend descriptors. The artifact section gives the exact source,
descriptor, and generated-output paths.

\begin{figure}[H]
\centering
\includegraphics[width=0.96\linewidth,height=\textheight,keepaspectratio,alt={Fail-closed lowering pipeline}]{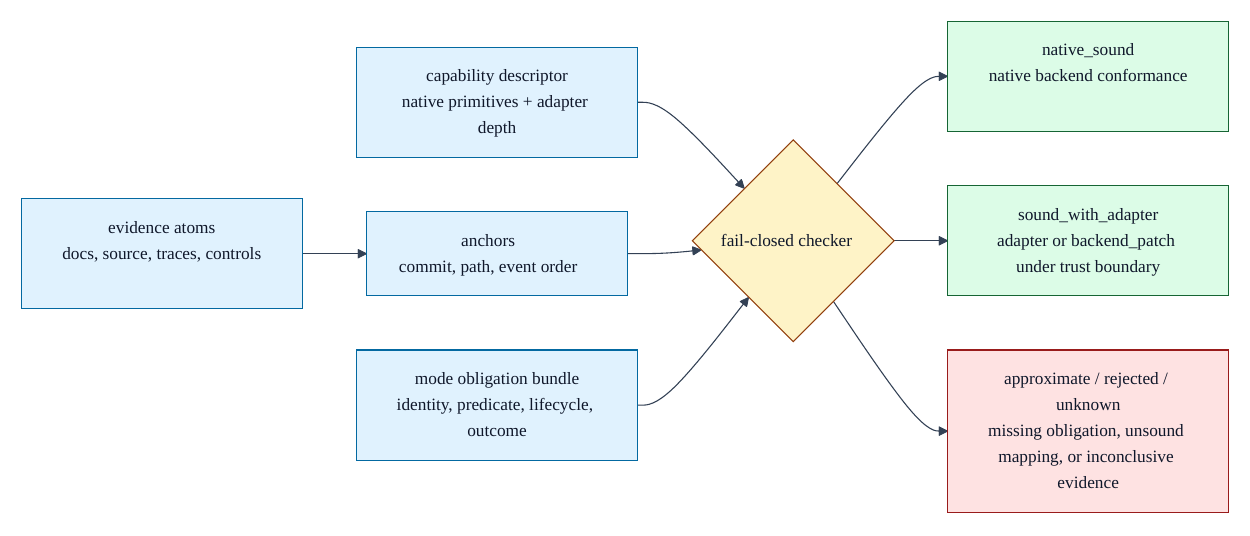}
\caption{Fail-closed lowering pipeline}
\end{figure}

The checker enforces seven rules:

\textbf{Table 2: Fail-closed checker rules.}

{\def\LTcaptype{none} 
\begin{longtable}[]{@{}
  >{\raggedright\arraybackslash}p{(\linewidth - 2\tabcolsep) * \real{0.5000}}
  >{\raggedright\arraybackslash}p{(\linewidth - 2\tabcolsep) * \real{0.5000}}@{}}
\toprule\noalign{}
\begin{minipage}[b]{\linewidth}\raggedright
Rule
\end{minipage} & \begin{minipage}[b]{\linewidth}\raggedright
Consequence
\end{minipage} \\
\midrule\noalign{}
\endhead
\bottomrule\noalign{}
\endlastfoot
Approximation signals never satisfy obligations by themselves. &
Priority, duration, block events, storage tiers, and routing can make a
row plausible without making it conformant. \\
Obligations are evidence-gated. & An obligation counts only when
\texttt{obligation\_\allowbreak{}evidence} is \texttt{supported} and has an
anchor. \\
Observed evidence atoms must be anchored. & Pressure atoms such as
\texttt{pressure\_\allowbreak{}controls\_\allowbreak{}observed} cannot upgrade a row without trace
anchors. \\
Evidence scope matters. & Docs-only and source-inspection adapter rows
do not become adapter-scoped positives without adequate runtime or
conformance evidence. \\
Adapter depth constrains obligations. & Telemetry joins can reconstruct
observations; they cannot create allocator, scheduler, routing, or
restoration enforcement. \\
Telemetry cannot create enforcement. & Block events can describe what
happened, but cannot create refusal, victim exclusion, blocking claim
ids, or restoration failure outcomes. \\
Ambiguity fails closed. & Missing registry, unstable ids, cache-identity
mismatch, partial event streams, duplicate ownership, and ambiguous
event order do not produce claim outcomes. \\
\end{longtable}
}

The generated matrix uses five labels:

\textbf{Table 3: Checker labels and interpretation in this paper.}

{\def\LTcaptype{none} 
\begin{longtable}[]{@{}
  >{\raggedright\arraybackslash}p{(\linewidth - 4\tabcolsep) * \real{0.3333}}
  >{\raggedright\arraybackslash}p{(\linewidth - 4\tabcolsep) * \real{0.3333}}
  >{\raggedright\arraybackslash}p{(\linewidth - 4\tabcolsep) * \real{0.3333}}@{}}
\toprule\noalign{}
\begin{minipage}[b]{\linewidth}\raggedright
Checker label
\end{minipage} & \begin{minipage}[b]{\linewidth}\raggedright
Meaning
\end{minipage} & \begin{minipage}[b]{\linewidth}\raggedright
Interpretation in this paper
\end{minipage} \\
\midrule\noalign{}
\endhead
\bottomrule\noalign{}
\endlastfoot
\texttt{native\_\allowbreak{}sound} & The backend natively satisfies all obligations
for the mode without an adapter or patch. & Native backend conformance.
No current descriptor has this label. \\
\texttt{sound\_\allowbreak{}with\_\allowbreak{}adapter} & A represented adapter, hook, or patch
supplies all obligations under explicit trust assumptions and adequate
evidence. & Use a precise subtype, such as adapter-observational
evidence, adapter-policy evidence under controlled pressure, or
backend-patch mechanism. \\
\texttt{approximate} & The backend exposes related primitives but misses
at least one required obligation. & Useful substrate, not ResidentClaim
conformance. \\
\texttt{rejected} & The proposed lowering would misrepresent the
contract and must fail closed. & Unsound mapping. \\
\texttt{unknown} & Current evidence is inconclusive or the relevant
primitive was not exercised. & No classification claim yet. \\
\end{longtable}
}

The label names are implementation vocabulary rather than a formal proof
claim. In particular, \texttt{native\_\allowbreak{}sound} and
\texttt{sound\_\allowbreak{}with\_\allowbreak{}adapter} mean that all required obligations are
supported by curated, anchored descriptor evidence at the stated depth.
They do not certify unaudited runtime paths.

The risky label is \texttt{sound\_\allowbreak{}with\_\allowbreak{}adapter}, so the manuscript uses
more precise phrases. Adapter-observational evidence reconstructs
selected claim-scoped observations. Adapter-policy evidence under
controlled pressure reconstructs selected policy behavior in the tested
pressure regime. A backend-patch mechanism supplies obligations inside a
modified runtime path. These are not interchangeable and none implies
native backend conformance.

\textbf{Descriptor-to-row examples.}

Positive row: TensorRT \texttt{soft\_\allowbreak{}priority\ +\ telemetry\_\allowbreak{}join}
combines pre-registered claims, stable joins, priority assignment,
pressure controls, and observed priority influence. For mode
\texttt{soft\_\allowbreak{}priority} at \texttt{telemetry\_\allowbreak{}join} depth, those fields
become adapter-policy evidence under controlled pressure. The checker
label is \texttt{sound\_\allowbreak{}with\_\allowbreak{}adapter}, but the row is positive only for
the tested controlled pressure family; it is not native TensorRT
conformance or hard protection.

Approximate row: TensorRT
\texttt{claim\_\allowbreak{}joined\_\allowbreak{}offload\ +\ generic\_\allowbreak{}onboard\_\allowbreak{}counters} has one
claim-joined offload plus generic onboard counters. For
\texttt{offloadable}, the row remains \texttt{approximate} because it
lacks claim-joined restore-before-reuse and a restoration-failure
outcome. Storage movement is useful substrate evidence, not offloadable
ResidentClaim conformance.

\section{5. Adapter Depths and Trust
Boundaries}\label{adapter-depths-and-trust-boundaries}

The adapter boundary is part of the result because an adapter is part of
the trusted computing base for any adapter-scoped row. The current depth
ladder is:

\textbf{Table 4: Adapter depths and trust boundaries.}

{\def\LTcaptype{none} 
\begin{longtable}[]{@{}
  >{\raggedright\arraybackslash}p{(\linewidth - 2\tabcolsep) * \real{0.5000}}
  >{\raggedright\arraybackslash}p{(\linewidth - 2\tabcolsep) * \real{0.5000}}@{}}
\toprule\noalign{}
\begin{minipage}[b]{\linewidth}\raggedright
Adapter depth
\end{minipage} & \begin{minipage}[b]{\linewidth}\raggedright
Meaning in this study
\end{minipage} \\
\midrule\noalign{}
\endhead
\bottomrule\noalign{}
\endlastfoot
\texttt{none} & Only native backend obligations count. \\
\texttt{telemetry\_\allowbreak{}join} & External registry and deterministic
request/cache/event joins reconstruct selected observations. \\
\texttt{claim\_\allowbreak{}registry} & Accepted claim state is represented outside
the backend or at a shallow integration point. \\
\texttt{storage\_\allowbreak{}restorability} & Storage/load-back hooks can report
claim-scoped restoration success and failure. \\
\texttt{routing\_\allowbreak{}hook} & Routing hooks can attribute route cost,
placement, and later reuse to claims. \\
\texttt{scheduler\_\allowbreak{}hook} & Scheduler hooks can emit an explicit
active/resident conflict action and report blocking claim ids. \\
\texttt{allocator\_\allowbreak{}hook} & Allocator hooks can exclude accepted claim
victims before violation. \\
\texttt{backend\_\allowbreak{}patch} & A modified backend path supplies the
represented obligations. \\
\end{longtable}
}

The \texttt{telemetry\_\allowbreak{}join} depth is intentionally narrow. It can
satisfy selected observational obligations only under these
preconditions:

\textbf{Table 5: Telemetry-join preconditions.}

{\def\LTcaptype{none} 
\begin{longtable}[]{@{}
  >{\raggedright\arraybackslash}p{(\linewidth - 2\tabcolsep) * \real{0.5000}}
  >{\raggedright\arraybackslash}p{(\linewidth - 2\tabcolsep) * \real{0.5000}}@{}}
\toprule\noalign{}
\begin{minipage}[b]{\linewidth}\raggedright
Preconditions for \texttt{telemetry\_\allowbreak{}join}
\end{minipage} & \begin{minipage}[b]{\linewidth}\raggedright
Required meaning
\end{minipage} \\
\midrule\noalign{}
\endhead
\bottomrule\noalign{}
\endlastfoot
External accepted-claim registry & Claims are pre-registered before the
events being classified. \\
Stable claim id & Claim identity is stable and distinct from backend
request ids. \\
Reusable object id & The reusable KV object is named consistently. \\
Fixed materialization predicate & The adapter knows the predicate being
evaluated, such as \texttt{leading\_\allowbreak{}prefix\_\allowbreak{}at\_\allowbreak{}least(k)}. \\
Deterministic request-token map & Request construction is derived
independently from tokenizer output. \\
Fixed cache identity & Joins are scoped to model, tokenizer/token-hash
domain, runtime, namespace, adapter identity, and block size. \\
Named observation point & Materialization is evaluated at a named event
or time boundary. \\
Joinable backend events & Backend events expose enough block hashes,
token ids, cache levels, and ordering to join to the predicate. \\
Ambiguity fails closed & Missing fields, duplicate owners, cache
mismatch, post-hoc claim naming, and event ambiguity produce
insufficient evidence. \\
\end{longtable}
}

Telemetry-only adapters cannot satisfy enforcement obligations. They
cannot create hard protection, victim exclusion, active conflict
actions, blocking claim ids, demotion-before-loss, expiry boundaries,
restoration failure outcomes, or claim harm. They can only make selected
observations defensible when the join keys are trustworthy.

The deeper adapter levels have analogous trust boundaries. They are
stronger than telemetry joins only when the descriptor anchors the
specific behavior claimed.

\textbf{Deeper adapter trust boundaries.}

{\def\LTcaptype{none} 
\begin{longtable}[]{@{}
  >{\raggedright\arraybackslash}p{(\linewidth - 6\tabcolsep) * \real{0.2500}}
  >{\raggedright\arraybackslash}p{(\linewidth - 6\tabcolsep) * \real{0.2500}}
  >{\raggedright\arraybackslash}p{(\linewidth - 6\tabcolsep) * \real{0.2500}}
  >{\raggedright\arraybackslash}p{(\linewidth - 6\tabcolsep) * \real{0.2500}}@{}}
\toprule\noalign{}
\begin{minipage}[b]{\linewidth}\raggedright
Depth
\end{minipage} & \begin{minipage}[b]{\linewidth}\raggedright
Preconditions for counting obligations
\end{minipage} & \begin{minipage}[b]{\linewidth}\raggedright
What can go wrong
\end{minipage} & \begin{minipage}[b]{\linewidth}\raggedright
Fail-closed handling
\end{minipage} \\
\midrule\noalign{}
\endhead
\bottomrule\noalign{}
\endlastfoot
\texttt{storage\_\allowbreak{}restorability} & Claim is accepted before offload;
store/load hooks name the same claim and cache identity; restoration is
ordered before predicate-satisfying reuse; failure outcome is emitted
before fallback or terminal handling. & Hook can lie about media, race
with reuse, lose load-failure events, or relabel fallback recompute as
restoration. & Missing order, wrong claim id, storage-only evidence,
fallback recompute, or generic counters do not satisfy
\texttt{offloadable}. \\
\texttt{routing\_\allowbreak{}hook} & Route decision sees the accepted claim;
worker/cache namespace is fixed; route cost, placement, and later reuse
are all attributed to the same claim and predicate. & Router can
optimize overlap without accepting responsibility, misattribute worker
state, or lose later hit/miss attribution. & Routing-only evidence
remains approximate until all route, placement, and reuse obligations
are claim scoped. \\
\texttt{scheduler\_\allowbreak{}hook} & Conflict is detected at a named
active/resident boundary; the claimed action is emitted before terminal
handling; blocking claim ids are stable and specific. & Scheduler can
emit post-hoc reasons, smear blocking ids across claims, race with
termination, or report refusal without the claimed resident cause. & The
checker requires evidence for whichever conflict action is claimed.
Ambiguous or smeared attribution fails closed. \\
\texttt{allocator\_\allowbreak{}hook} & Allocator exposes the victim candidate set,
protected resident footprint, and ordered exclusion or release before
predicate-breaking loss. & Allocator can hide victims, race with
compaction, report aggregate pressure only, or lose protected-victim
identity. & Hard-protection rows remain rejected without victim
exclusion or an explicit alternative contract transition. \\
\texttt{backend\_\allowbreak{}patch} & Patch stack and imported source are
identified; patched code lies on the exercised runtime path; adjacent
hooks not used by the claim are absent or disabled; analyzer checks
order and identity after the run. & Patch can bypass the real path, rely
on adjacent instrumentation, misattribute claims, or overstate coverage
beyond the exercised branch. & Patch evidence is a backend-patch witness
only, not native support; the row counts only anchored obligations
supplied by the patch and exercised trace. \\
\end{longtable}
}

Adapter composition is conjunctive and fail-closed. Selecting a deeper
adapter depth permits obligations from shallower represented adapters
only when each obligation has anchored evidence at a depth allowed to
supply it. For \texttt{hard\_\allowbreak{}protected},
\texttt{scheduler\_\allowbreak{}hook\ +\ allocator\_\allowbreak{}hook} means both sides are
required: scheduler evidence may supply the explicit conflict action and
blocking claim ids, while allocator evidence must separately supply
victim exclusion before violation. Telemetry joins cannot fill either
enforcement role, and a scheduler-only or allocator-only row remains
rejected unless an explicit alternative contract transition supplies the
missing obligation.

These trust assumptions are part of the result. An adapter can lie,
race, drop events, or misattribute claims. The checker cannot prove such
failures absent; it handles them by refusing to upgrade ambiguous
evidence into a claim outcome.

\section{6. External Runtime Boundary
Studies}\label{external-runtime-boundary-studies}

The central result is a semantic lowering table, not a feature
compatibility chart. It summarizes the generated matrix and the boundary
memos without treating feature names as conformance.

\textbf{Table 6: Runtime boundary study summary.}

{\def\LTcaptype{none} 
\begin{longtable}[]{@{}
  >{\raggedright\arraybackslash}p{(\linewidth - 4\tabcolsep) * \real{0.3333}}
  >{\raggedright\arraybackslash}p{(\linewidth - 4\tabcolsep) * \real{0.3333}}
  >{\raggedright\arraybackslash}p{(\linewidth - 4\tabcolsep) * \real{0.3333}}@{}}
\toprule\noalign{}
\begin{minipage}[b]{\linewidth}\raggedright
Substrate
\end{minipage} & \begin{minipage}[b]{\linewidth}\raggedright
Best current evidence
\end{minipage} & \begin{minipage}[b]{\linewidth}\raggedright
Fail-closed boundary
\end{minipage} \\
\midrule\noalign{}
\endhead
\bottomrule\noalign{}
\endlastfoot
Patched vLLM connector/scheduler boundary & \texttt{backend\_\allowbreak{}patch}
witness for the \texttt{offloadable} lifecycle/outcome bundle. & Not
upstream/native vLLM support, not production offload performance, and
not pre-admission scheduler refusal. \\
TensorRT-LLM & Adapter-observational \texttt{best\_\allowbreak{}effort} and
adapter-policy \texttt{soft\_\allowbreak{}priority} under controlled pressure. &
Active no-evict and storage movement do not supply future-resident hard
protection or offloadable outcomes. \\
SGLang/HiCache & Adapter-observational \texttt{best\_\allowbreak{}effort} under
explicit telemetry-join preconditions. & Storage tiers and cached-token
telemetry do not supply claim-scoped restoration success/failure. \\
Dynamo-style routing & Docs-backed boundary evidence and routing
comparator. & KV-aware routing alone lacks claim identity, route cost,
placement, and later reuse attribution. \\
\end{longtable}
}

\subsection{6.1 TensorRT-LLM}\label{tensorrt-llm}

TensorRT-LLM is the strongest public-runtime boundary study in this
artifact set. The audited source descriptor at commit
\texttt{06cff70502} records priority, duration, KV events, host
secondary cache, and active no-evict mechanisms as substrates. The
executable container descriptors represent the official TensorRT-LLM
1.3.0rc14 and 1.3.0rc15 release containers separately, with runtime
traces for block-event telemetry, priority/duration assignment,
controlled soft-priority pressure, duration boundary probing, active
no-evict, and corrected adaptive block-tier movement.

The positive TensorRT rows are adapter-scoped. The
\texttt{best\_\allowbreak{}effort\ +\ telemetry\_\allowbreak{}join} row reconstructs
materialization for selected leading-prefix claims from an external
registry, tokenizer-derived request map, cache identity, named
observation point, and joinable block/token events. The
\texttt{soft\_\allowbreak{}priority\ +\ telemetry\_\allowbreak{}join} row is adapter-policy
evidence under controlled pressure: original trials lost lower-priority
prompt B first in 5/5 runs, swapped trials lost lower-priority prompt A
first in 5/5 runs, equal-priority trials lost both tracked prompts in
the same first-loss event in 3/3 runs, both claims were joinable before
pressure in 13/13 trials, no pre-pressure claim loss occurred in 13/13
trials, and materialization changes were reconstructable in 16/16
tracked loss claims.

The negative rows are equally important. \texttt{GUARANTEED\_\allowbreak{}NO\_\allowbreak{}EVICT}
is available and runnable, but the observed API/source scope is
active-request scheduling: the trace emitted 103 KV events, all
\texttt{created} or \texttt{stored}, and did not expose native
future-resident claim identity, explicit conflict action, blocking claim
ids, harm fields, or victim exclusion before violation. The checker
therefore rejects native \texttt{hard\_\allowbreak{}protected} and
\texttt{hard\_\allowbreak{}protected\ +\ telemetry\_\allowbreak{}join}.

Duration remains approximation. A duration boundary trace assigned a 15s
short claim and a 60s long control, then applied pressure after the
short duration and before the long duration. It observed 480 pressure
events, 448 stored events, 32 removed events, and 995 removed block
hashes, but no duration/expiry event fields and no joined loss for
either duration claim. That is retention metadata, not a claim-scoped
expiry boundary.

Offload remains below the \texttt{offloadable} bar for public TensorRT
evidence. A corrected May 25, 2026 TensorRT-LLM 1.3.0rc15 adaptive
capture observed same-prompt block-hash tier movement: tracked
assignment hashes moved from cache level 0 to 1 under pressure and back
from 1 to 0 on before-removal same-prompt reuse. The corrected
no-retention control shows this movement does not require retention
config on either assignment or reuse. Different-prompt reuse did not
onboard the tracked hashes, and post-removal same-prompt reuse
stored/recomputed rather than restoring. This is runtime-observed
block-tier movement substrate evidence, not ResidentClaim offloadable
conformance: TensorRT still lacks accepted claim identity, a
claim-scoped materialization predicate, restoration-failure outcome, and
active/resident outcome attribution.

\subsection{6.2 SGLang/HiCache}\label{sglanghicache}

SGLang/HiCache is the calibrated storage comparator. Source inspection
at \texttt{bbe9c7e} shows radix priority, lock/reference state,
protected and evictable accounting, GPU/host/storage tiers, prefetch,
write-back, load-back, storage success/failure flags, cached-token
details, and block events. These are strong mechanisms, but most are
request, page, hash, token, or cache-node scoped rather than
claim-outcome scoped.

The positive SGLang row is \texttt{best\_\allowbreak{}effort\ +\ telemetry\_\allowbreak{}join}. A
model-level trace and fresh-process repeat used an external
accepted-claim registry, a tokenizer-derived request-token map, cache
identity, request cached-token telemetry, and native block events to
reconstruct materialization for \texttt{leading\_\allowbreak{}prefix\_\allowbreak{}at\_\allowbreak{}least(17)}.
The cold shared-prefix request reported \texttt{cached\_\allowbreak{}tokens=0}, the
warm shared-prefix request reported \texttt{cached\_\allowbreak{}tokens=35}, the
salted wrong-prefix control reported \texttt{cached\_\allowbreak{}tokens=0}, and
native event replay returned 100 \texttt{Block\allowbreak{}Stored} events. This is
adapter-observational evidence, not native SGLang ResidentClaim support.

The HiCache/storage row remains approximate. HiCache exposes real
storage and load-back mechanics, but the current evidence does not
report claim-scoped restoration success before reuse, restoration
failure as refusal/demotion/expiry or harm, ordered offload lifecycle,
or claim harm attribution. A guarded
\texttt{cached\_\allowbreak{}tokens\_\allowbreak{}details.\allowbreak{}storage} proxy can support
storage-backed materialization observations, but its ceiling explicitly
excludes restoration success/failure semantics.

\subsection{6.3 Dynamo-Style Routing}\label{dynamo-style-routing}

Dynamo-style KV routing is included as a first-class boundary row
because routing and placement are natural places to preserve future
reuse. The public docs-backed descriptor records KV-aware routing,
overlap and prefill-cost routing, worker KV event tracking, queue
priority, and placement decisions as useful orchestration substrates.
The \texttt{routed\_\allowbreak{}reuse} mode asks for more: claim-scoped route cost,
placement attribution, later reuse-routing attribution, materialization
predicate, accepted claim identity, and claim-scoped telemetry. Those
obligations are not shown in the current docs-backed evidence, so
routing remains approximate rather than conformant.

Dynamo would upgrade from routing comparator to ResidentClaim evidence
only if it exposed claim-scoped route cost, placement attribution, and
later reuse success/failure attribution for an accepted claim.

\section{7. Patched vLLM Connector/Scheduler-Boundary
Mechanism}\label{patched-vllm-connectorscheduler-boundary-mechanism}

The local patched vLLM connector/scheduler-boundary mechanism supplies
the missing \texttt{offloadable} lifecycle/outcome witness at
\texttt{backend\_\allowbreak{}patch} depth. It patches the vLLM pydev
\texttt{Offloading\allowbreak{}Connector} path rather than replacing the offload path
with a standalone simulator. Native vLLM supplies real in-process
connector lookup, store/load job creation, worker transfer
submission/completion, and failed-load propagation into the scheduler
invalid-block path. The patch supplies ResidentClaim metadata, lifecycle
joins, scheduler-boundary restoration failure telemetry, and
claim-scoped outcome emission.

\begin{figure}[H]
\centering
\includegraphics[width=0.96\linewidth,height=\textheight,keepaspectratio,alt={Patched vLLM connector/scheduler-boundary witness}]{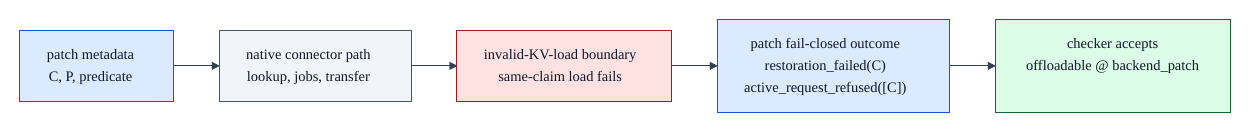}
\caption{Patched vLLM connector/scheduler-boundary witness}
\end{figure}

The mechanism has three trust boundaries. Native vLLM supplies connector
lookup, store/load job creation, worker transfer submission/completion,
and failed-load propagation into scheduler invalid-block handling. The
patch supplies the accepted claim metadata, lifecycle joins, controlled
same-claim failure injection, scheduler-boundary restoration failure
telemetry, and active outcome emission. The artifact analyzer supplies
no runtime behavior; it only checks order, claim match, and controls
after the run. This separation is why the row is a backend-patch witness
rather than native support.

In this mechanism, ``accepted'' means that pre-request claim metadata is
accepted by the patched path before the relevant lifecycle events. It
does not mean an upstream user-facing vLLM ResidentClaim API exists.
``Refused'' means the patched scheduler emits a fail-closed active
outcome after same-claim restoration failure is detected at the
invalid-KV-load path and before terminal error handling. The
scheduler-side active-refusal event is not pre-admission scheduling
refusal.

``Scheduler-boundary'' means the patched path emits the active outcome
at vLLM's invalid-KV-load handling boundary before terminal
request-finished/error handling. It is not initial scheduling admission
and not upstream/native scheduler API support.

Failure injection is disabled unless the resident-claim load-failure
flag is enabled. When enabled, the hook matches only
\texttt{CPU\ -\textgreater{}\ GPU} transfers and can filter by claim id.
Unclaimed generic failures require a separate flag. The decisive
positive path is not ``a request failed''; it is the ordered sequence:
accepted claim metadata exists, the claim has been offloaded, reuse
requires restoration, the matching \texttt{CPU\ -\textgreater{}\ GPU}
load fails under controlled injection, scheduler invalid-block handling
identifies the affected request, emits
\texttt{scheduler\_\allowbreak{}resident\_\allowbreak{}claim\_\allowbreak{}restoration\_\allowbreak{}failed}, and emits
\texttt{scheduler\_\allowbreak{}active\_\allowbreak{}request\_\allowbreak{}refused} with that claim in
\texttt{blocking\_\allowbreak{}claim\_\allowbreak{}ids} before the request-finished boundary. The
previous connector-level \texttt{resident\_\allowbreak{}claim\_\allowbreak{}restoration\_\allowbreak{}failed}
and \texttt{active\_\allowbreak{}request\_\allowbreak{}refused} events remain as downstream
evidence.

The witness tables use short aliases for layout; these are exact
artifact event names and the observed failure reason:

\begin{Shaded}
\begin{Highlighting}[]
\NormalTok{E0  request\_initialized}
\NormalTok{E1  offload\_lookup\_result}
\NormalTok{E2  offload\_store\_job\_created}
\NormalTok{E3  offload\_worker\_transfer\_submitted}
\NormalTok{E4  offload\_worker\_transfer\_finished}
\NormalTok{E5  resident\_claim\_offloaded}
\NormalTok{E6  resident\_claim\_restore\_required}
\NormalTok{E7  offload\_load\_job\_created}
\NormalTok{E8  resident\_claim\_restored}
\NormalTok{E9  offload\_job\_completed}
\NormalTok{E10 offload\_request\_finished\_no\_pending\_jobs}
\NormalTok{E11 offload\_worker\_load\_failed}
\NormalTok{E12 scheduler\_resident\_claim\_restoration\_failed}
\NormalTok{E13 scheduler\_active\_request\_refused}
\NormalTok{E14 offload\_request\_finished\_pending\_jobs}
\NormalTok{F0  controlled\_resident\_claim\_cpu\_to\_gpu\_load\_failure}
\end{Highlighting}
\end{Shaded}

\textbf{vLLM witness path A: successful offload/load observation.}

{\def\LTcaptype{none} 
\begin{longtable}[]{@{}
  >{\raggedleft\arraybackslash}p{(\linewidth - 4\tabcolsep) * \real{0.4000}}
  >{\raggedright\arraybackslash}p{(\linewidth - 4\tabcolsep) * \real{0.3000}}
  >{\raggedright\arraybackslash}p{(\linewidth - 4\tabcolsep) * \real{0.3000}}@{}}
\toprule\noalign{}
\begin{minipage}[b]{\linewidth}\raggedleft
Order
\end{minipage} & \begin{minipage}[b]{\linewidth}\raggedright
Events
\end{minipage} & \begin{minipage}[b]{\linewidth}\raggedright
What it establishes
\end{minipage} \\
\midrule\noalign{}
\endhead
\bottomrule\noalign{}
\endlastfoot
1 & E0 & Claim metadata exists before the offload/load lifecycle. \\
2 & E2, E3, E4 success, E5 & Store/offload job and GPU-to-CPU transfer
complete for the claim. \\
3 & E0 reuse, E1 hit=448 & Reuse lookup hits the offloaded claim
footprint. \\
4 & E6, E7 & Restore/load is required before reuse can satisfy the
predicate. \\
5 & E3, E4 success, E8, E9 & CPU-to-GPU restoration succeeds before
reuse completion. \\
6 & E10 plus summary \texttt{served} & Reuse is served after
predicate-satisfying cached tokens are restored. \\
\end{longtable}
}

\textbf{vLLM witness path B: same-claim failure outcome.}

{\def\LTcaptype{none} 
\begin{longtable}[]{@{}
  >{\raggedleft\arraybackslash}p{(\linewidth - 4\tabcolsep) * \real{0.4000}}
  >{\raggedright\arraybackslash}p{(\linewidth - 4\tabcolsep) * \real{0.3000}}
  >{\raggedright\arraybackslash}p{(\linewidth - 4\tabcolsep) * \real{0.3000}}@{}}
\toprule\noalign{}
\begin{minipage}[b]{\linewidth}\raggedleft
Order
\end{minipage} & \begin{minipage}[b]{\linewidth}\raggedright
Events
\end{minipage} & \begin{minipage}[b]{\linewidth}\raggedright
What it establishes
\end{minipage} \\
\midrule\noalign{}
\endhead
\bottomrule\noalign{}
\endlastfoot
1 & E0 & Claim metadata exists before lifecycle events. \\
2 & E2, E3, E4 success, E5 & The same claim was offloaded. \\
3 & E1 hit=448, E6 & Reuse lookup hits and restoration is required. \\
4 & E7, E3, E4 failure F0 & The CPU-to-GPU load failure is tied to the
same claim. \\
5 & E11 & vLLM's invalid-KV-load path has affected block evidence. \\
6 & E12 with matching claim and \texttt{FINISHED\_\allowbreak{}ERROR} &
Scheduler-boundary failure is claim-scoped. \\
7 & E13 with \texttt{blocking\_claim\_ids={[}C{]}} & The active request
is fail-closed because of the blocking claim. \\
8 & E14 after E12/E13 & The scheduler outcome occurs before terminal
request-finished handling. \\
\end{longtable}
}

\textbf{vLLM witness path C: multi-claim attribution control.}

The clean attribution rerun creates two accepted ResidentClaim-shaped
identities in one process. The non-target claim restores successfully
through the offload/load path. The target claim arms controlled
\texttt{CPU\ -\textgreater{}\ GPU} restoration failure and receives
scheduler-boundary
\texttt{scheduler\_\allowbreak{}resident\_\allowbreak{}claim\_\allowbreak{}restoration\_\allowbreak{}failed},
\texttt{scheduler\_\allowbreak{}active\_\allowbreak{}request\_\allowbreak{}refused}, and
\texttt{blocking\_\allowbreak{}claim\_\allowbreak{}ids} attribution. In 3/3 repetitions, scheduler
events and \texttt{blocking\_\allowbreak{}claim\_\allowbreak{}ids} named only the target claim;
the non-target claim restored successfully in 3/3 repetitions and
received 0/3 failure or refusal attributions. This is a
target-attribution control for the backend-patch witness. It is not
upstream/native vLLM support and not pre-admission refusal.

The rerun provenance is deliberately narrow. The attribution summary is
recorded in the paper artifact submodule at commit
\texttt{558a069b\allowbreak{}d6fd897f\allowbreak{}4c5e7273\allowbreak{}e95d05f6\allowbreak{}9c835a23}. The vLLM source tree
was at \texttt{0b129d97\allowbreak{}135c1f01\allowbreak{}cc907008\allowbreak{}b7c93ad6\allowbreak{}5256ea64} with no
uncommitted changes, and the imported vLLM module matched that source in
3/3 repetitions. The adjacent \texttt{0004} prefix-cache materialization
hook emitted 0/3 events in the raw traces. The target-attribution result
therefore depends on the clean \texttt{0001}--\texttt{0003} connector
and scheduler-boundary stack, not on the adjacent prefix-cache hook.

Equivalent event names would be acceptable in another runtime. What
matters is the anchored order: same accepted claim, offload,
restore-required reuse, restore-before-reuse on success, and same-claim
restoration failure becoming a claim-scoped fail-closed active outcome
on failure.

This mechanism is not native/upstream vLLM support. It is not
scheduler-native pre-admission refusal, and it is not a production
offload performance result. Its purpose is narrower and important: it
demonstrates that the missing offload lifecycle/outcome obligations are
implementable in a real connector path with scheduler-boundary
invalid-load handling.

\textbf{What would make public runtimes native?}

{\def\LTcaptype{none} 
\begin{longtable}[]{@{}
  >{\raggedright\arraybackslash}p{(\linewidth - 2\tabcolsep) * \real{0.5000}}
  >{\raggedright\arraybackslash}p{(\linewidth - 2\tabcolsep) * \real{0.5000}}@{}}
\toprule\noalign{}
\begin{minipage}[b]{\linewidth}\raggedright
Runtime family
\end{minipage} & \begin{minipage}[b]{\linewidth}\raggedright
Native evidence the checker would accept
\end{minipage} \\
\midrule\noalign{}
\endhead
\bottomrule\noalign{}
\endlastfoot
TensorRT-LLM and SGLang/HiCache & Accepted claim identity, a
materialization predicate, ordered lifecycle events, and claim-scoped
outcomes. For offload, the trace must also expose restore-before-reuse
and a restoration-failure outcome. \\
Dynamo-style routing & Accepted claim identity, a materialization
predicate, claim-scoped route cost, placement attribution, and later
reuse success/failure attribution. \\
Any public runtime & Equivalent anchored evidence is enough; the checker
requires obligation conformance, not the patch's event names. \\
\end{longtable}
}

\section{8. Evaluation}\label{evaluation}

The evaluation answers four questions. The first two are matrix
questions: does the checker reject false positives, and do studied
runtimes natively satisfy the obligations under current public evidence?
The second two are mechanism questions: can the missing offloadable
semantics be implemented in a real runtime path, and is the mechanism
stable in the tested setting with characterized artifact/analyzer costs
but no production performance claim?

\textbf{Table 7: Evaluation questions and answers.}

{\def\LTcaptype{none} 
\begin{longtable}[]{@{}
  >{\raggedright\arraybackslash}p{(\linewidth - 4\tabcolsep) * \real{0.3333}}
  >{\raggedright\arraybackslash}p{(\linewidth - 4\tabcolsep) * \real{0.3333}}
  >{\raggedright\arraybackslash}p{(\linewidth - 4\tabcolsep) * \real{0.3333}}@{}}
\toprule\noalign{}
\begin{minipage}[b]{\linewidth}\raggedright
Question
\end{minipage} & \begin{minipage}[b]{\linewidth}\raggedright
Evidence used
\end{minipage} & \begin{minipage}[b]{\linewidth}\raggedright
Answer
\end{minipage} \\
\midrule\noalign{}
\endhead
\bottomrule\noalign{}
\endlastfoot
Does the checker reject false positives? & Generated matrix,
bad-lowering counterexamples, descriptor provenance, independent
TensorRT descriptor audit, and descriptor/evidence mutation controls. &
Yes. Feature-only mappings and small evidence mutations remain
approximate, rejected, unknown, or invalid unless the required
obligations and observed atoms are present. \\
Do studied runtimes natively satisfy the obligations under current
public evidence? & TensorRT-LLM, SGLang/HiCache, and Dynamo descriptors
plus boundary studies. & No native conformance is established. Public
evidence supports substrates and selected adapter positives. \\
Can the missing offloadable semantics be implemented in a real runtime
path? & Patched vLLM connector/scheduler-boundary run. & Yes, at
\texttt{backend\_\allowbreak{}patch} depth: real in-process offload/load behavior
carries claim metadata and same-claim restoration failure becomes a
scheduler-boundary fail-closed active outcome. \\
Is the mechanism stable in the tested setting? & 131 repeated subprocess
runs, 3 multi-claim attribution repetitions, and the common analyzer. &
Yes for the local semantic gate: 131/131 subprocesses completed with
valid event sequences, and 3/3 multi-claim repetitions attributed
failure/refusal only to the target claim. \\
\end{longtable}
}

\subsection{8.1 Matrix Evaluation}\label{matrix-evaluation}

The generated matrix contains no \texttt{native\_\allowbreak{}sound} rows. This is
expected. TensorRT-LLM and SGLang/HiCache produce selected adapter
positives and many approximation rows. Dynamo routing produces
approximation rows for routing/placement. The patched vLLM reference
produces backend-patch witness rows only at \texttt{backend\_\allowbreak{}patch}
depth for represented state-machine and connector obligations.

The matrix is regenerated from manually curated descriptors and mode
obligations, not edited by hand. The bad-lowering suite separately
checks feature-table false positives against the same checker.

The positive rows also have an explicit provenance index. The generated
companion files \texttt{results/\allowbreak{}descriptor-\allowbreak{}provenance.\allowbreak{}*} in the
artifact, mirrored as
\texttt{studies/\allowbreak{}lowering/\allowbreak{}results/\allowbreak{}descriptor-\allowbreak{}provenance.\allowbreak{}*} in this
worktree, keep the full per-obligation anchor list. The compact version
uses obligation codes so each row can be traced to the obligation
bundle: \texttt{I=claim\_\allowbreak{}identity}, \texttt{A=explicit\_\allowbreak{}acceptance},
\texttt{P=materialization\_\allowbreak{}predicate}, \texttt{F=footprint\_\allowbreak{}accounting},
\texttt{L=ordered\_\allowbreak{}lifecycle\_\allowbreak{}events},
\texttt{M=claim\_\allowbreak{}materialized\_\allowbreak{}event},
\texttt{D=claim\_\allowbreak{}demoted\_\allowbreak{}before\_\allowbreak{}loss},
\texttt{E=claim\_\allowbreak{}expired\_\allowbreak{}boundary}, \texttt{R=offload\_\allowbreak{}restorability},
\texttt{RF=restoration\_\allowbreak{}failure\_\allowbreak{}outcome},
\texttt{V=victim\_\allowbreak{}exclusion\_\allowbreak{}before\_\allowbreak{}violation},
\texttt{X=explicit\_\allowbreak{}conflict\_\allowbreak{}action}, \texttt{B=blocking\_\allowbreak{}claim\_\allowbreak{}ids},
\texttt{H=claim\_\allowbreak{}harm\_\allowbreak{}attribution},
\texttt{T=claim\_\allowbreak{}scoped\_\allowbreak{}telemetry}, and
\texttt{Pr=priority\_\allowbreak{}influence}.

\textbf{Descriptor provenance for positive rows.}

{\def\LTcaptype{none} 
\begin{longtable}[]{@{}
  >{\raggedright\arraybackslash}p{(\linewidth - 8\tabcolsep) * \real{0.2000}}
  >{\raggedright\arraybackslash}p{(\linewidth - 8\tabcolsep) * \real{0.2000}}
  >{\raggedright\arraybackslash}p{(\linewidth - 8\tabcolsep) * \real{0.2000}}
  >{\raggedright\arraybackslash}p{(\linewidth - 8\tabcolsep) * \real{0.2000}}
  >{\raggedright\arraybackslash}p{(\linewidth - 8\tabcolsep) * \real{0.2000}}@{}}
\toprule\noalign{}
\begin{minipage}[b]{\linewidth}\raggedright
Descriptor path
\end{minipage} & \begin{minipage}[b]{\linewidth}\raggedright
Mode / depth / evidence
\end{minipage} & \begin{minipage}[b]{\linewidth}\raggedright
Anchor/result path
\end{minipage} & \begin{minipage}[b]{\linewidth}\raggedright
Obligations
\end{minipage} & \begin{minipage}[b]{\linewidth}\raggedright
Non-claim
\end{minipage} \\
\midrule\noalign{}
\endhead
\bottomrule\noalign{}
\endlastfoot
\texttt{capabilities/\allowbreak{}sglang\_\allowbreak{}hicache\_\allowbreak{}bbe9c7e.\allowbreak{}yaml} &
\texttt{best\_\allowbreak{}effort} / \texttt{telemetry\_\allowbreak{}join} /
\texttt{litmus\_\allowbreak{}trace} &
\texttt{results/\allowbreak{}sglang/\allowbreak{}sglang\_\allowbreak{}claim\_\allowbreak{}telemetry\_\allowbreak{}adapter\_\allowbreak{}join\_\allowbreak{}summary.\allowbreak{}json};
fresh repeat & I, P, M, T & No native SGLang; no offloadable restoration
outcome. \\
\texttt{capabilities/\allowbreak{}tensorrt\_\allowbreak{}llm\_\allowbreak{}1\_\allowbreak{}3\_\allowbreak{}0rc14\_\allowbreak{}container.\allowbreak{}yaml} &
\texttt{best\_\allowbreak{}effort} / \texttt{telemetry\_\allowbreak{}join} /
\texttt{litmus\_\allowbreak{}trace} &
\texttt{adapters/\allowbreak{}tests/\allowbreak{}test\_\allowbreak{}telemetry\_\allowbreak{}join.\allowbreak{}py}; TensorRT L3
block-event summary & I, P, M, T & No native TensorRT; no
hard/offloadable claim. \\
\texttt{capabilities/\allowbreak{}tensorrt\_\allowbreak{}llm\_\allowbreak{}1\_\allowbreak{}3\_\allowbreak{}0rc14\_\allowbreak{}container.\allowbreak{}yaml} &
\texttt{soft\_\allowbreak{}priority} / \texttt{telemetry\_\allowbreak{}join} /
\texttt{litmus\_\allowbreak{}trace} &
\texttt{results/\allowbreak{}tensorrt/\allowbreak{}trt\_\allowbreak{}soft\_\allowbreak{}priority\_\allowbreak{}pressure\_\allowbreak{}controls\_\allowbreak{}summary.\allowbreak{}json};
claim-join summary & I, Pr, T & Controlled-pressure adapter-policy row
only. \\
\texttt{capabilities/\allowbreak{}vllm\_\allowbreak{}patched.\allowbreak{}yaml} & \texttt{best\_\allowbreak{}effort} /
\texttt{backend\_\allowbreak{}patch} / \texttt{conformance\_\allowbreak{}trace} &
\texttt{artifacts/\allowbreak{}kv-\allowbreak{}residency-\allowbreak{}vllm-\allowbreak{}arbiter/\allowbreak{}artifacts/\allowbreak{}conformance/\allowbreak{}results.\allowbreak{}json}
& I, P, M, T & Patch witness; no upstream/native vLLM. \\
\texttt{capabilities/\allowbreak{}vllm\_\allowbreak{}patched.\allowbreak{}yaml} & \texttt{demotable} /
\texttt{backend\_\allowbreak{}patch} / \texttt{conformance\_\allowbreak{}trace} &
\texttt{artifacts/\allowbreak{}kv-\allowbreak{}residency-\allowbreak{}vllm-\allowbreak{}arbiter/\allowbreak{}artifacts/\allowbreak{}conformance/\allowbreak{}results.\allowbreak{}json}
& I, A, D, L & Patch witness; no production performance. \\
\texttt{capabilities/\allowbreak{}vllm\_\allowbreak{}patched.\allowbreak{}yaml} & \texttt{expiring} /
\texttt{backend\_\allowbreak{}patch} / \texttt{conformance\_\allowbreak{}trace} &
\texttt{artifacts/\allowbreak{}kv-\allowbreak{}residency-\allowbreak{}vllm-\allowbreak{}arbiter/\allowbreak{}artifacts/\allowbreak{}conformance/\allowbreak{}results.\allowbreak{}json}
& I, A, E, L & Patch witness; no upstream/native vLLM. \\
\texttt{capabilities/\allowbreak{}vllm\_\allowbreak{}patched.\allowbreak{}yaml} & \texttt{hard\_\allowbreak{}protected} /
\texttt{backend\_\allowbreak{}patch} / \texttt{conformance\_\allowbreak{}trace} &
\texttt{artifacts/\allowbreak{}kv-\allowbreak{}residency-\allowbreak{}vllm-\allowbreak{}arbiter/\allowbreak{}artifacts/\allowbreak{}conformance/\allowbreak{}results.\allowbreak{}json}
& I, A, P, F, V, X, B, H, L & Patch witness; not native admission. \\
\texttt{capabilities/\allowbreak{}vllm\_\allowbreak{}patched.\allowbreak{}yaml} & \texttt{offloadable} /
\texttt{backend\_\allowbreak{}patch} / \texttt{conformance\_\allowbreak{}trace} &
\texttt{results/\allowbreak{}vllm\_\allowbreak{}scheduler\_\allowbreak{}boundary/\allowbreak{}repetitions/\allowbreak{}20260522Tresident\_\allowbreak{}claim\_\allowbreak{}scheduler\_\allowbreak{}boundary/\allowbreak{}aggregate.\allowbreak{}json};
\texttt{results/\allowbreak{}vllm-\allowbreak{}multi-\allowbreak{}claim-\allowbreak{}attribution-\allowbreak{}control.\allowbreak{}json} & I, A, P, R,
RF, L, H & Connector patch witness; no production/upstream claim. \\
\end{longtable}
}

As an audit control, the TensorRT rc14 descriptor has a
reviewer-auditable independent descriptor pass at
\texttt{results/\allowbreak{}tensorrt-\allowbreak{}rc14-\allowbreak{}independent-\allowbreak{}descriptor-\allowbreak{}audit.\allowbreak{}*}. That pass
re-derives all 14 TensorRT rc14 rows from the descriptor's anchored
obligation evidence, evidence atoms, mode obligations, adapter-depth
rules, and preconditions without reading
\texttt{results/\allowbreak{}lowering-\allowbreak{}matrix.\allowbreak{}*} as the answer. It agrees with the
descriptor checker rows in 14/14 cases, including the two positive
adapter rows and the rejected \texttt{hard\_\allowbreak{}protected} rows. This is an
independent audit over curated evidence, not proof that TensorRT-LLM
behavior is complete.

\subsection{8.2 Descriptor Mutation
Controls}\label{descriptor-mutation-controls}

The descriptor/evidence mutation controls address the self-certification
objection directly. They mutate positive descriptor rows or replay
near-miss runtime summaries and require fail-closed behavior. The
generated controls pass in 16/16 cases.

The mutations cover anchor deletion or empty anchors, supported evidence
changed to partial/unknown/missing, observed pressure atoms without
anchors, adapter scope weakened to docs-only or source-inspection,
missing telemetry-join preconditions, routing-only evidence, wrong claim
id failure attribution, post-hoc claim naming, restore-after-reuse
ordering, fallback recompute, generic counters, storage-only offload
evidence, and routing-only evidence.

This establishes a narrow property: the checker and runtime gates are
sensitive to small evidence, identity, order, precondition,
adapter-scope, and source-shape mutations that would otherwise create
false positives. It does not prove that descriptors are independently
complete. Descriptors remain curated evidence summaries, and unaudited
runtime behavior remains outside the proof boundary.

\subsection{8.3 Connector Repetition
Results}\label{connector-repetition-results}

The repeated connector evaluation ran the artifact harness in
subprocesses against the local patched vLLM pydev source. It used the
real connector path in small in-process runs. The artifact section lists
the canonical artifact aggregate and manifest.

\textbf{Table 8: Connector repetition semantic gates.}

{\def\LTcaptype{none} 
\begin{longtable}[]{@{}
  >{\raggedright\arraybackslash}p{(\linewidth - 6\tabcolsep) * \real{0.2308}}
  >{\raggedleft\arraybackslash}p{(\linewidth - 6\tabcolsep) * \real{0.3077}}
  >{\raggedright\arraybackslash}p{(\linewidth - 6\tabcolsep) * \real{0.2308}}
  >{\raggedright\arraybackslash}p{(\linewidth - 6\tabcolsep) * \real{0.2308}}@{}}
\toprule\noalign{}
\begin{minipage}[b]{\linewidth}\raggedright
Gate
\end{minipage} & \begin{minipage}[b]{\linewidth}\raggedleft
Runs
\end{minipage} & \begin{minipage}[b]{\linewidth}\raggedright
Result
\end{minipage} & \begin{minipage}[b]{\linewidth}\raggedright
Interpretation
\end{minipage} \\
\midrule\noalign{}
\endhead
\bottomrule\noalign{}
\endlastfoot
Event-sequence validity & 131 & 131/131 valid & All subprocess runs
produced analyzer-parseable event order. \\
Positive observation path & 30 & 30/30 observation passes & Claim
metadata flows through successful offload/load behavior. \\
Same-claim failure outcome & 30 & 30/30 failure-outcome passes &
Controlled \texttt{CPU\ -\textgreater{}\ GPU} load failure for the
accepted claim becomes scheduler-boundary fail-closed active outcome. \\
False-positive controls & 41 & 0/41 failure-outcome passes & Ordinary
offload without claim, unclaimed failure, wrong-claim failure, fallback
recompute, and generic counters do not satisfy the gate. \\
\end{longtable}
}

Timing, byte, and analyzer-latency numbers are artifact diagnostics
rather than semantic gates; they are summarized in Appendix A.

The clean multi-claim rerun adds attribution evidence that the broader
repetition suite did not isolate. The target claim fails controlled
\texttt{CPU\ -\textgreater{}\ GPU} restoration and receives
scheduler-boundary restoration-failed/refused/blocking ids. The
non-target claim restores successfully and receives no failure/refusal
attribution. The gate passes in 3/3 repetitions. This is a backend-patch
witness for target attribution, not upstream/native vLLM and not
pre-admission refusal.

\subsection{8.4 Interpretation}\label{interpretation}

The repeated-run result is a stable semantic gate result. The same
analyzer that accepts same-claim controlled failure rejects missing
events, success-only observations as failure evidence, ordinary offload
without ResidentClaim metadata, wrong claim ids, unclaimed failures,
fallback recomputation, and generic counters. For
\texttt{claimed\_\allowbreak{}load\_\allowbreak{}failure}, all 30 repeated rows record the
scheduler-side failure outcome, scheduler-side refusal, same-claim
match, event-before-termination order, connector-level outcome, and no
native scheduler admission refusal. That is the important property: the
mechanism is falsifiable at the obligation boundary.

\section{9. False Positive
Counterexamples}\label{false-positive-counterexamples}

The bad-lowering suite records feature-table inferences that a less
strict study might accidentally call supported. Each case is checked
against the same obligation relation as the main matrix.

\textbf{Table 9: False-positive counterexamples.}

{\def\LTcaptype{none} 
\begin{longtable}[]{@{}
  >{\raggedright\arraybackslash}p{(\linewidth - 4\tabcolsep) * \real{0.3333}}
  >{\raggedright\arraybackslash}p{(\linewidth - 4\tabcolsep) * \real{0.3333}}
  >{\raggedright\arraybackslash}p{(\linewidth - 4\tabcolsep) * \real{0.3333}}@{}}
\toprule\noalign{}
\begin{minipage}[b]{\linewidth}\raggedright
Naive inference
\end{minipage} & \begin{minipage}[b]{\linewidth}\raggedright
Checker result
\end{minipage} & \begin{minipage}[b]{\linewidth}\raggedright
Why it fails
\end{minipage} \\
\midrule\noalign{}
\endhead
\bottomrule\noalign{}
\endlastfoot
\texttt{priority\_value\_in\_event\ -\textgreater{}\ soft\_priority} &
\texttt{approximate} & A priority value is block metadata unless
priority influence and claim-scoped telemetry are both established. \\
\texttt{active\_no\_evict\ -\textgreater{}\ future\_resident\ hard\_protected}
& \texttt{rejected} & Active no-evict can protect running requests
without accepted future-resident claim identity, victim exclusion,
explicit conflict action, blocking claim ids, or harm attribution. \\
\texttt{duration\_metadata\ -\textgreater{}\ expiring} &
\texttt{approximate} & Duration metadata does not report the
claim-scoped boundary where responsibility ends. \\
\texttt{storage\_tier\ -\textgreater{}\ offloadable} &
\texttt{approximate} & Storage movement does not show restoration before
reuse or claim-scoped restoration failure. \\
\texttt{claim\_joined\_offload\ +\ generic\_onboard\_counters\ -\textgreater{}\ offloadable}
& \texttt{approximate} & Even a claim-joined offload plus generic
onboard counters does not establish claim-joined restore-before-reuse or
a restoration-failure outcome. \\
\texttt{same\_prompt\_block\_tier\_movement\ -\textgreater{}\ offloadable}
& \texttt{approximate} & Corrected TensorRT rc15 rows observed
same-prompt tracked hashes moving 0 -\textgreater{} 1 under pressure and
1 -\textgreater{} 0 before removal without retention config, but exposed
no native claim identity, predicate, failure outcome, lifecycle, or
harm/refusal/demotion/expiry attribution. \\
\texttt{kv\_aware\_routing\ -\textgreater{}\ routed\_reuse} &
\texttt{approximate} & Routing needs route cost, placement, and future
reuse success/failure attributed to an accepted claim. \\
\texttt{block\_removed\ -\textgreater{}\ claim\_harm} & invalid lowering
claim & Removed blocks are ordinary cache behavior unless accepted claim
identity, predicate-breaking loss, and claim harm attribution are
present. \\
fallback recompute after failed load -\textgreater{} restored
offloadable claim & rejected by connector gate & Recomputing after a
failed load is not evidence that the accepted offloaded claim was
restored. \\
wrong-claim or unclaimed load failure -\textgreater{} restoration
failure outcome & rejected by connector gate & The failure must be tied
to the same accepted claim; generic or wrong-claim failures are not
claim outcomes. \\
\end{longtable}
}

These counterexamples are central because they explain why negative rows
are not merely missing evidence. Some lowerings would actively
misrepresent the contract and must fail closed.

\section{10. Limitations and Threats to
Validity}\label{limitations-and-threats-to-validity}

\textbf{Table 10: Limitations and consequences.}

{\def\LTcaptype{none} 
\begin{longtable}[]{@{}
  >{\raggedright\arraybackslash}p{(\linewidth - 2\tabcolsep) * \real{0.5000}}
  >{\raggedright\arraybackslash}p{(\linewidth - 2\tabcolsep) * \real{0.5000}}@{}}
\toprule\noalign{}
\begin{minipage}[b]{\linewidth}\raggedright
Limitation
\end{minipage} & \begin{minipage}[b]{\linewidth}\raggedright
Consequence
\end{minipage} \\
\midrule\noalign{}
\endhead
\bottomrule\noalign{}
\endlastfoot
No native conformance is shown for public TensorRT-LLM, SGLang/HiCache,
Dynamo, or upstream vLLM evidence. & The positive claims are
adapter-scoped or patch-scoped, not native backend support. \\
The patched vLLM connector result is local \texttt{backend\_\allowbreak{}patch}
evidence. & It demonstrates implementability of the missing
lifecycle/outcome mechanism, not upstream support. \\
Refusal is scheduler-boundary at the invalid-KV-load handling boundary,
not pre-admission. & The paper does not claim upstream/native scheduler
admission/refusal. \\
Restoration failure uses controlled injection. & The gate tests ordered
same-claim failure handling, not arbitrary production failure
coverage. \\
The connector runs are small, in-process, and single-GPU. & The results
support semantic gate stability, not broad model/GPU/concurrency
generality. \\
TTFT and vLLM stat-logger connector metrics are unavailable. &
\texttt{disable\_\allowbreak{}log\_\allowbreak{}stats=True} is required because of an
OffloadingConnector metrics serialization issue. \\
TensorRT runtime traces use specific container paths and pressure
families. & Do not generalize to every TensorRT version, backend, model,
or workload. \\
SGLang \texttt{best\_\allowbreak{}effort} evidence is model-level adapter evidence. &
It does not upgrade HiCache storage to \texttt{offloadable} or native
SGLang conformance. \\
Dynamo evidence is docs-backed. & It is a routing/placement boundary
row, not a runtime conformance trace. \\
Generated historical artifacts contain absolute local paths as
provenance. & Those fields are not portable commands and should be read
as historical run metadata. \\
Local vLLM patch content is provided as format patches. & The patch is
not an upstream branch and should be applied deliberately to a clean
matching base. \\
\end{longtable}
}

The most important non-claims are:

\textbf{Table 11: Calibrated non-claims.}

{\def\LTcaptype{none} 
\begin{longtable}[]{@{}
  >{\raggedright\arraybackslash}p{(\linewidth - 2\tabcolsep) * \real{0.5000}}
  >{\raggedright\arraybackslash}p{(\linewidth - 2\tabcolsep) * \real{0.5000}}@{}}
\toprule\noalign{}
\begin{minipage}[b]{\linewidth}\raggedright
Do not claim
\end{minipage} & \begin{minipage}[b]{\linewidth}\raggedright
Calibrated claim
\end{minipage} \\
\midrule\noalign{}
\endhead
\bottomrule\noalign{}
\endlastfoot
Upstream/native vLLM ResidentClaim support. & A local patched vLLM
connector mechanism supplies selected lifecycle/outcome obligations at
\texttt{backend\_\allowbreak{}patch} depth. \\
Native ResidentClaim conformance for TensorRT-LLM, SGLang/HiCache, or
Dynamo. & They expose strong substrates and selected
adapter-observational or adapter-policy positives. No native conformance
is shown. \\
Host cache, storage tiers, or generic counters implement
\texttt{offloadable}. & Offloadable requires restoration before reuse
and a claim-scoped restoration-failure outcome. \\
\texttt{GUARANTEED\_\allowbreak{}NO\_\allowbreak{}EVICT} implements future-resident
\texttt{hard\_\allowbreak{}protected}. & The observed TensorRT no-evict path is
active-request scoped. \\
Production readiness or general serving-overhead bounds for the
connector result. & It is a small controlled semantic mechanism
evaluation. \\
The vLLM result is upstream/native scheduler admission. & It is a local
scheduler-boundary patch at invalid-KV-load handling, not pre-admission
or upstream support. \\
Routing implements \texttt{routed\_\allowbreak{}reuse}. & Routing remains an
approximation until route cost, placement, and reuse are
claim-attributed. \\
\end{longtable}
}

\section{11. Related Work and Prior-Art
Boundary}\label{related-work-and-prior-art-boundary}

TensorRT-LLM is the closest primitive-level comparator. Its versioned
KV-cache documentation describes cross-request reuse, prioritized LRU,
retention priority/duration fields, and secondary-memory offload
(\href{https://github.com/NVIDIA/TensorRT-LLM/blob/06cff70502/docs/source/features/kvcache.md}{NVIDIA
TensorRT-LLM KV Cache System, commit \texttt{06cff70502}}). This paper
treats those mechanisms as serious substrates. The boundary is that the
public evidence does not expose accepted ResidentClaim identity,
claim-scoped expiry, restoration-before-reuse, restoration-failure
outcomes, or active-side conflict outcomes.

SGLang/HiCache is the strongest storage/offload comparator. The audited
HiCache documentation describes RadixAttention-derived prefix reuse,
GPU/host/L3 tiers, local match, L3 prefetch, write-back policies,
CPU-to-GPU transfer optimizations, and L3 backends including Mooncake
and LMCache
(\href{https://github.com/sgl-project/sglang/blob/bbe9c7e/docs/advanced_features/hicache_design.md}{SGLang
HiCache design, commit \texttt{bbe9c7e}}). Those mechanisms move toward
useful future reuse, but request/page/hash/token scoped storage
mechanics are not, by themselves, claim-scoped restoration lifecycle
outcomes.

Dynamo-style KV-aware routing is the orchestration comparator. NVIDIA's
Dynamo router documentation describes KV routing modes that use worker
KV events, cache-overlap accounting, queue policy, prefill-cost
modeling, and placement decisions
(\href{https://docs.nvidia.com/dynamo/user-guides/kv-cache-aware-routing}{NVIDIA
Dynamo Router Guide}). The \texttt{routed\_\allowbreak{}reuse} obligation bundle asks
for a different boundary: route cost, placement, and later reuse must be
attributed to an accepted claim and materialization predicate. Current
public docs evidence supports a routing substrate, not ResidentClaim
routed-reuse conformance.

vLLM supplies the implementation substrate used by the positive witness.
PagedAttention introduced block-structured KV memory management for LLM
serving (\href{https://doi.org/10.1145/3600006.3613165}{Kwon et al.,
SOSP 2023}), and vLLM documents automatic prefix caching for shared
prefixes
(\href{https://docs.vllm.ai/en/v0.10.1/features/automatic_prefix_caching.html}{vLLM
APC documentation, 2025-08-07}). The public vLLM offloading connector
material describes asynchronous connector load/store behavior and
CPU-backed offload
(\href{https://vllm.ai/blog/2026-01-08-kv-offloading-connector}{vLLM
offloading connector blog, 2026-01-08};
\href{https://github.com/vllm-project/vllm/issues/19854}{vLLM RFC
\#19854}). The local mechanism in this paper uses those kinds of
connector/scheduler surfaces as substrate while adding ResidentClaim
identity and outcomes in a patch. A separate vLLM retention RFC confirms
that priority/TTL-style retention is an active serving-community design
topic, not a solved ResidentClaim contract
(\href{https://github.com/vllm-project/vllm/issues/37003}{vLLM RFC
\#37003}).

Adjacent systems sharpen the non-claim. KVCache Cache in the Wild
characterizes production reuse patterns and workload-aware KV eviction
(\href{https://doi.org/10.48550/arXiv.2506.02634}{Wang et al.,
arXiv:2506.02634}). Continuum uses TTL-based KV retention for multi-turn
agent scheduling (\href{https://doi.org/10.48550/arXiv.2511.02230}{Li et
al., arXiv:2511.02230}). KVFlow uses workflow structure for prefix
caching and prefetch in multi-agent workflows
(\href{https://doi.org/10.48550/arXiv.2507.07400}{Pan et al.,
arXiv:2507.07400}). Marconi studies prefix caching for hybrid LLMs with
reuse/cost-aware cache decisions
(\href{https://doi.org/10.48550/arXiv.2411.19379}{Pan et al.,
arXiv:2411.19379}). Pie exposes programmable serving control, including
KV-cache strategies, to application code
(\href{https://doi.org/10.48550/arXiv.2510.24051}{Gim et al.,
arXiv:2510.24051}). Mooncake and LMCache make KV cache storage and
transfer first-class system surfaces
(\href{https://doi.org/10.1145/3773772}{Qin et al., ACM TOS 2025};
\href{https://doi.org/10.48550/arXiv.2510.09665}{Liu et al.,
arXiv:2510.09665}). Newer adjacent systems such as TokenCake, FlowKV,
Tutti, and FlexKV further strengthen the non-claim boundary because they
provide KV movement/scheduling/offload mechanisms; the distinction here
is accepted-claim lifecycle and outcome conformance
(\href{https://doi.org/10.48550/arXiv.2510.18586}{Bian et al.,
arXiv:2510.18586}; \href{https://doi.org/10.48550/arXiv.2504.03775}{Li
et al., arXiv:2504.03775};
\href{https://doi.org/10.48550/arXiv.2605.03375}{Qiu et al.,
arXiv:2605.03375};
\href{https://docs.nvidia.com/dynamo/integrations/flex-kv}{NVIDIA FlexKV
documentation}). These systems constrain broad novelty claims about
future-reuse policy, workflow-aware cache management, and distributed KV
storage. They are compatible with the narrower contribution here: a
fail-closed conformance relation for accepted future-KV obligations and
a patched systems witness for one missing offload lifecycle/outcome
path.

\section{12. Artifact Availability and Reproducibility
Notes}\label{artifact-availability-and-reproducibility-notes}

The audit surface for this paper is the curated artifact repository
\texttt{resident-\allowbreak{}kv-\allowbreak{}lowering-\allowbreak{}artifact} at commit
\texttt{b9f82f45\allowbreak{}6e56e484\allowbreak{}54a9b4e0\allowbreak{}c608c2c7\allowbreak{}83d0cbdb}:

\begin{Shaded}
\begin{Highlighting}[]
\NormalTok{https://github.com/gustavgauge/resident{-}kv{-}lowering{-}artifact.git}
\end{Highlighting}
\end{Shaded}

The curated snapshot contains the checker, capability descriptors,
generated matrix, bad-lowering counterexamples, selected runtime
summaries, vLLM scheduler-boundary evidence, descriptor/evidence
mutation controls, the multi-claim attribution control summary, and
format patches. The key artifact paths are:

\begin{itemize}
\tightlist
\item
  \texttt{checker/\allowbreak{}}, \texttt{checker/\allowbreak{}modes.\allowbreak{}yaml}, and
  \texttt{checker/\allowbreak{}generate\_\allowbreak{}matrix.\allowbreak{}py}
\item
  \texttt{capabilities/\allowbreak{}} for backend descriptors
\item
  \texttt{results/\allowbreak{}lowering-\allowbreak{}matrix.\allowbreak{}md} and
  \texttt{results/\allowbreak{}lowering-\allowbreak{}matrix.\allowbreak{}json}
\item
  \texttt{results/\allowbreak{}central-\allowbreak{}result-\allowbreak{}table.\allowbreak{}md}
\item
  \texttt{bad\_\allowbreak{}lowering/\allowbreak{}} and
  \texttt{results/\allowbreak{}bad-\allowbreak{}lowering-\allowbreak{}counterexamples.\allowbreak{}*}
\item
  \texttt{checker/\allowbreak{}generate\_\allowbreak{}mutation\_\allowbreak{}controls.\allowbreak{}py} and
  \texttt{results/\allowbreak{}descriptor-\allowbreak{}evidence-\allowbreak{}mutation-\allowbreak{}controls.\allowbreak{}*}
\item
  \texttt{checker/\allowbreak{}generate\_\allowbreak{}descriptor\_\allowbreak{}provenance.\allowbreak{}py},
  \texttt{results/\allowbreak{}descriptor-\allowbreak{}provenance.\allowbreak{}*}, and
  \texttt{results/\allowbreak{}tensorrt-\allowbreak{}rc14-\allowbreak{}independent-\allowbreak{}descriptor-\allowbreak{}audit.\allowbreak{}*}
\item
  \texttt{results/\allowbreak{}vllm-\allowbreak{}multi-\allowbreak{}claim-\allowbreak{}attribution-\allowbreak{}control.\allowbreak{}*}
\item
  \texttt{results/\allowbreak{}vllm\_\allowbreak{}scheduler\_\allowbreak{}boundary/\allowbreak{}} for aggregate
  scheduler-boundary evidence
\item
  \texttt{results/\allowbreak{}tensorrt/\allowbreak{}tensorrt\_\allowbreak{}rc15\_\allowbreak{}corrected\_\allowbreak{}adaptive\_\allowbreak{}restore/\allowbreak{}}
  for the corrected TensorRT rc15 retention-confound rerun and
  mechanical gate
\item
  \texttt{vllm\_\allowbreak{}patches/\allowbreak{}} for patches against vLLM base
  \texttt{9b9d5dba\allowbreak{}ab852a1c\allowbreak{}615fe83a\allowbreak{}7f92881d\allowbreak{}353503db}
\end{itemize}

The curated snapshot also records two provenance identities:

\begin{itemize}
\tightlist
\item
  arbiter branch \texttt{resident-\allowbreak{}claim-\allowbreak{}lifecycle-\allowbreak{}outcome} at commit
  prefix \texttt{f2658451}
\item
  local vLLM patches through commit prefix \texttt{0b129d97}
\end{itemize}

The artifact README records the full hashes. Generated historical
evidence may preserve absolute local paths as provenance; those fields
are not portable commands.

The strengthened connector attribution evidence used in this revision is
summarized in the public artifact as
\texttt{results/\allowbreak{}vllm-\allowbreak{}multi-\allowbreak{}claim-\allowbreak{}attribution-\allowbreak{}control.\allowbreak{}*}. The raw rerun
provenance is the local paper artifact submodule
\texttt{artifacts/\allowbreak{}kv-\allowbreak{}residency-\allowbreak{}vllm-\allowbreak{}arbiter} at commit
\texttt{558a069b\allowbreak{}d6fd897f\allowbreak{}4c5e7273\allowbreak{}e95d05f6\allowbreak{}9c835a23}. That rerun uses a
clean vLLM \texttt{0001}--\texttt{0003} patch stack at source head
\texttt{0b129d97\allowbreak{}135c1f01\allowbreak{}cc907008\allowbreak{}b7c93ad6\allowbreak{}5256ea64}; the adjacent
\texttt{0004} prefix-cache materialization hook is absent from the raw
traces.

In the curated artifact repository, use the README commands:

\begin{Shaded}
\begin{Highlighting}[]
\ExtensionTok{uv}\NormalTok{ run }\AttributeTok{{-}{-}with}\NormalTok{ pytest }\AttributeTok{{-}{-}with}\NormalTok{ pyyaml pytest }\AttributeTok{{-}q}\NormalTok{ checker/tests bad\_lowering/tests adapters/tests}
\ExtensionTok{uv}\NormalTok{ run }\AttributeTok{{-}{-}with}\NormalTok{ pyyaml python checker/generate\_matrix.py}
\ExtensionTok{uv}\NormalTok{ run }\AttributeTok{{-}{-}with}\NormalTok{ pyyaml python checker/generate\_mutation\_controls.py}
\ExtensionTok{uv}\NormalTok{ run }\AttributeTok{{-}{-}with}\NormalTok{ pyyaml python checker/generate\_descriptor\_provenance.py}
\ExtensionTok{uv}\NormalTok{ run }\AttributeTok{{-}{-}with}\NormalTok{ pyyaml python bad\_lowering/check\_bad\_lowerings.py}
\end{Highlighting}
\end{Shaded}

\section{Appendix A. Artifact
Diagnostics}\label{appendix-a.-artifact-diagnostics}

The semantic gates in the main evaluation are event order, claim
identity, failure/refusal attribution, and false-positive rejection. The
repeated connector artifacts also record timing, byte, and analyzer-cost
diagnostics so the evidence scale is inspectable. These diagnostics are
not serving-overhead, TTFT, throughput, cache-hit, or
production-performance results.

For the \texttt{claimed\_\allowbreak{}load\_\allowbreak{}failure} row in the 30-run
scheduler-boundary suite, median/p95 local diagnostics are: resident
wall time 0.187152/0.204366 s, reuse wall time 0.003159/0.003468 s,
event size 26,702/26,702 bytes, analyzer runtime 128,560/146,290 ns,
failure-to-outcome latency 151,544/174,749 ns, and
\texttt{restoration\_\allowbreak{}failed} to \texttt{active\_\allowbreak{}request\_\allowbreak{}refused}
latency 80,319.5/92,189 ns. The 3-run multi-claim attribution control
records 46,619/46,620 event bytes, 168,610/171,470 analyzer ns,
143,790/144,809 failure-to-outcome ns, and 74,560/74,760
\texttt{restoration\_\allowbreak{}failed} to \texttt{active\_\allowbreak{}request\_\allowbreak{}refused} ns.

TTFT is unavailable because the connector runs use
\texttt{disable\_\allowbreak{}log\_\allowbreak{}stats=True} to avoid a vLLM
\texttt{Offloading\allowbreak{}Connector} metrics serialization assertion: the
offload stats path stores \texttt{Offloading\allowbreak{}Operation\allowbreak{}Metrics} objects
while the reducer and Prometheus observer expect serialized
dictionaries. The lower wall time in some success-path rows is not a
speedup result.

\section{13. Conclusion}\label{conclusion}

ResidentClaim lowering is an obligation problem, not a feature-name
problem. TensorRT-LLM, SGLang/HiCache, Dynamo-style routing, and vLLM
connector paths all expose useful KV mechanisms, but useful mechanisms
do not automatically become accepted future-KV obligations. The
fail-closed checker and boundary studies show how the artifact makes
that distinction mechanically for manually curated, anchored
descriptors: adapter-scoped observations can be positive under explicit
preconditions, storage and routing can remain useful substrates without
becoming conformance, and unsound feature-table lowerings can be
rejected.

The calibrated systems result is the local patched vLLM
connector/scheduler-boundary mechanism. It carries claim metadata
through a real connector path and reports same-claim restoration failure
as a scheduler-boundary claim-scoped outcome followed by fail-closed
active refusal at the invalid-KV-load branch under controlled load
failure. That is enough to demonstrate implementability of the missing
offload lifecycle/outcome semantics at patch depth, while preserving the
boundary that public runtime primitives, by themselves, do not establish
native ResidentClaim conformance.

\section{14. References}\label{references}

\begin{itemize}
\tightlist
\item
  Woosuk Kwon et al.~``Efficient Memory Management for Large Language
  Model Serving with PagedAttention.'' SOSP 2023. DOI:
  \url{https://doi.org/10.1145/3600006.3613165}.
\item
  NVIDIA. ``TensorRT-LLM KV Cache System.'' Versioned documentation at
  commit \texttt{06cff70502}, accessed 2026-05-23.
  \url{https://github.com/NVIDIA/TensorRT-LLM/blob/06cff70502/docs/source/features/kvcache.md}.
\item
  SGLang Project. ``HiCache System Design and Optimization.'' Versioned
  documentation at commit \texttt{bbe9c7e}, accessed 2026-05-23.
  \url{https://github.com/sgl-project/sglang/blob/bbe9c7e/docs/advanced_features/hicache_design.md}.
\item
  NVIDIA. ``Dynamo Router Guide.'' Documentation, accessed 2026-05-23.
  \url{https://docs.nvidia.com/dynamo/user-guides/kv-cache-aware-routing}.
\item
  vLLM Project. ``Automatic Prefix Caching.'' Documentation dated
  2025-08-07, accessed 2026-05-23.
  \url{https://docs.vllm.ai/en/v0.10.1/features/automatic_prefix_caching.html}.
\item
  vLLM Project. ``Inside vLLM's New KV Offloading Connector: Smarter
  Memory Transfer for Maximizing Inference Throughput.'' Blog post dated
  2026-01-08, accessed 2026-05-23.
  \url{https://vllm.ai/blog/2026-01-08-kv-offloading-connector}.
\item
  vLLM Project. ``{[}RFC{]}: KV cache offloading.'' GitHub issue
  \#19854, opened 2025-06-19, accessed 2026-05-23.
  \url{https://github.com/vllm-project/vllm/issues/19854}.
\item
  vLLM Project. ``{[}RFC{]}: Context-Aware KV-Cache Retention API
  (Prioritized Evictions).'' GitHub issue \#37003, opened 2026-03-13,
  accessed 2026-05-23.
  \url{https://github.com/vllm-project/vllm/issues/37003}.
\item
  Jiahao Wang et al.~``KVCache Cache in the Wild: Characterizing and
  Optimizing KVCache Cache at a Large Cloud Provider.''
  arXiv:2506.02634, 2025.
  \url{https://doi.org/10.48550/arXiv.2506.02634}.
\item
  Hanchen Li et al.~``Continuum: Efficient and Robust Multi-Turn LLM
  Agent Scheduling with KV Cache Time-to-Live.'' arXiv:2511.02230, 2025.
  \url{https://doi.org/10.48550/arXiv.2511.02230}.
\item
  Zaifeng Pan et al.~``KVFlow: Efficient Prefix Caching for Accelerating
  LLM-Based Multi-Agent Workflows.'' arXiv:2507.07400, 2025.
  \url{https://doi.org/10.48550/arXiv.2507.07400}.
\item
  Rui Pan et al.~``Marconi: Prefix Caching for the Era of Hybrid LLMs.''
  arXiv:2411.19379, 2024.
  \url{https://doi.org/10.48550/arXiv.2411.19379}.
\item
  In Gim et al.~``Pie: A Programmable Serving System for Emerging LLM
  Applications.'' arXiv:2510.24051, 2025.
  \url{https://doi.org/10.48550/arXiv.2510.24051}.
\item
  Ruoyu Qin et al.~``Mooncake: A KVCache-centric Disaggregated
  Architecture for LLM Serving.'' ACM Transactions on Storage, 2025.
  \url{https://doi.org/10.1145/3773772}.
\item
  Yuhan Liu et al.~``LMCache: An Efficient KV Cache Layer for
  Enterprise-Scale LLM Inference.'' arXiv:2510.09665, 2025.
  \url{https://doi.org/10.48550/arXiv.2510.09665}.
\item
  Zhuohang Bian et al.~``TokenCake: A KV-Cache-centric Serving Framework
  for LLM-based Multi-Agent Applications.'' arXiv:2510.18586, 2025.
  \url{https://doi.org/10.48550/arXiv.2510.18586}.
\item
  Weiqing Li et al.~``FlowKV: A Disaggregated Inference Framework with
  Low-Latency KV Cache Transfer and Load-Aware Scheduling.''
  arXiv:2504.03775,

  \begin{enumerate}
  \def\labelenumi{\arabic{enumi}.}
  \setcounter{enumi}{2024}
  \tightlist
  \item
    \url{https://doi.org/10.48550/arXiv.2504.03775}.
  \end{enumerate}
\item
  Shi Qiu et al.~``Tutti: Making SSD-Backed KV Cache Practical for
  Long-Context LLM Serving.'' arXiv:2605.03375, 2026.
  \url{https://doi.org/10.48550/arXiv.2605.03375}.
\item
  NVIDIA. ``FlexKV.'' Dynamo documentation, accessed 2026-05-23.
  \url{https://docs.nvidia.com/dynamo/integrations/flex-kv}.
\end{itemize}

\end{document}